\documentclass[aps,prd,
twocolumn,%
a4paper,superscriptaddress,nofootinbib%
]{revtex4}

\usepackage{graphicx}
\graphicspath{{Figures/Corner_new}}
\usepackage{amsmath}
\usepackage{multirow}
\usepackage{diagbox}

\usepackage[normalem]{ulem}

\usepackage{acronym}
\usepackage[
    colorlinks,
    linkcolor=blue,
    citecolor=blue,
    urlcolor=blue
]{hyperref}

\newcommand{\be}{\begin{equation}}
\newcommand{\ee}{\end{equation}}
\newcommand{\bea}{\begin{eqnarray}}
\newcommand{\eea}{\end{eqnarray}}
\newcommand{\bsub}{\begin{subequations}}
\newcommand{\esub}{\end{subequations}}
\newcommand{\nn}{\nonumber}
\newcommand{\ui}{\mathrm{i}}

\newcommand{\adress}{MOE Key Laboratory of TianQin Mission, TianQin Research Center for Gravitational Physics $\&$ School of Physics and Astronomy, Frontiers Science Center for TianQin, Gravitational Wave Research Center of CNSA, Sun Yat-sen University (Zhuhai Campus), Zhuhai 519082,  People's Republic of China}

\def\params{\boldsymbol{\theta}}
\def\data{\textbf{d}}
\def\noise{\textbf{n}}

\newcommand{\phd}{\texttt{IMRPhenomD} }

\newcommand{\tdh}{\tilde{h}}
\newcommand{\tdy}{\tilde{y}}
\newcommand{\tdA}{\tilde{A}}
\newcommand{\tdE}{\tilde{E}}
\newcommand{\tdT}{\tilde{T}}
\newcommand{\tdX}{\tilde{X}}
\newcommand{\tdY}{\tilde{Y}}
\newcommand{\tdZ}{\tilde{Z}}
\newcommand{\bfepsilon}{\boldsymbol{\epsilon}}
\newcommand{\bftheta}{\boldsymbol{\theta}}

\newcommand{\bfh}{\boldsymbol{h}}
\newcommand{\bfk}{\boldsymbol{k}}
\newcommand{\bfn}{\boldsymbol{n}}
\newcommand{\bfp}{\boldsymbol{p}}
\newcommand{\bfu}{\boldsymbol{u}}
\newcommand{\bfv}{\boldsymbol{v}}
\newcommand{\bfP}{\boldsymbol{P}}
\newcommand{\tdbfh}{\tilde{\boldsymbol{h}}}
\newcommand{\Mc}{\mathcal{M}_c}
\newcommand{\Tsr}{\mathcal{T}_{sr}}

\def\i{{\rm i}}
\def\mSun{{\rm M_\odot}}
\def\ccr#1{{\color{red} #1}}

\begin{document}
\title{Bayesian parameter estimation of massive black hole binaries with TianQin-LISA}

\author{Jie Gao}
\affiliation{\adress}

\author{Yi-Ming Hu}
\email{huyiming@sysu.edu.cn}
\affiliation{\adress}

\author{En-Kun Li}
\email{lienk@mail.sysu.edu.cn}
\affiliation{\adress}

\author{Jian-dong Zhang}
\affiliation{\adress}

\author{Jianwei Mei}
\affiliation{\adress}

\date{\today}

\begin{abstract}
    This paper analyses the impact of various parameter changes on the estimation of parameters for massive black hole binary (MBHB) systems using a Bayesian inference technique. 
    Several designed MBHB systems were chosen for comparison with a fiducial system to explore the influence of parameters such as sky location, inclination angle, anti-spin, large mass ratio and light mass. 
    And the two reported MBHB candidates named OJ287 and Tick-Tock are also considered.
    The study found that the network of TianQin and LISA can break certain degeneracies among different parameters, improving the estimation of parameters, particularly for extrinsic parameters. 
    Meanwhile, the degeneracies between different intrinsic parameters are highly sensitive to the value of the parameters.
    Additionally, the small inclination angles and limited detection of the inspiral phase can introduce significant bias in the estimation of parameters. 
    The presence of instrument noise will also introduce bias and worsen the precision.
    The paper concludes that the network of TianQin and LISA can significantly improve the estimation of extrinsic parameters by about one order of magnitude while yielding slight improvements in the intrinsic parameters. 
    Moreover, parameter estimation can still be subject to biases even with a sufficiently high signal-to-noise ratio if the detected signal does not encompass all stages of the inspiral, merger, and ringdown.
\end{abstract}

\maketitle

\acrodef{gw}[GW]{Gravitational Wave}
\acrodef{mbh}[MBH]{Massive Black Hole}
\acrodef{mbhb}[MBHB]{Massive Black Hole Binary}
\acrodef{agn}[AGN]{active galactic nuclei}
\acrodef{em}[EM]{Electromagnetic}
\acrodef{pta}[PTA]{pulsar timing array}
\acrodef{sgwb}[SGWB]{stochastic gravitational wave background}
\acrodef{mcmc}[MCMC]{Markov-chain Monte Carlo}
\acrodef{pn}[PN]{post-Newtonian}
\acrodef{nr}[NR]{numerical relativity}
\acrodef{tdi}[TDI]{time delay interferometry}
\acrodef{psd}[PSD]{power spectral density}
\acrodef{snr}[SNR]{Signal-to-Noise Ratio}
\acrodef{fim}[FIM]{Fisher Information Matrix}

\section{Introduction}
\label{sec:intro}

\acp{mbh}, with masses ranging from millions to tens of billions of solar masses, are widely believed to exist at the centers of almost all galaxies \cite{Lynden-Bell1969, Kormendy1995, Miyoshi1995, Tremmel2018}, including in the Galactic center \cite{Abuter2018} and the nucleus of the nearby elliptical galaxy M87  \cite{Collaboration2019}.
They are also found in a small number of low-mass dwarf galaxies \cite{Reines:2011na, Reines:2013pia, Baldassare:2019yua}. 
Given that massive galaxies often experience multiple mergers throughout their evolutionary history, it is natural to expect the presence of \acp{mbhb} in merging or merged galaxies \cite{Begelman:1980vb, 1989ApJ...343...47V, Komossa:2015cha}.
Observation methods based on \ac{em} radiation, such as the detection of semi-periodic signatures \cite{Graham2015, Graham2015a, Liu2015, Charisi2016, Chen2022} or double-peaked emission lines \cite{Tsalmantza2011, Ju2013}, have identified several potential candidates for \acp{mbhb} \cite{Liu:2014mga, Graham:2015gma, Bon:2016jtk, Charisi:2016fqw, Zhu:2020ybm, Koss:2023bvr}.
Of particular interest are two examples: the extensively studied blazar OJ\,287 \cite{Sillanpaa1988,Lehto1996,Valtonen2008} (OJ287), located at a redshift of $z = 0.306$;
and the Seyfert 1-type galaxy SDSS\,J143016.05 + 230344.4, which hosts a candidate \ac{mbhb} \cite{Oh2015, Jiang:2022aek} at redshift of $z=0.08105$ \cite{Oh:2011zv} (referred to `Tick-Tock' hereafter).

However, \ac{em} observations provide mostly indirect detection of \ac{mbhb} systems and are limited to accessing the most luminous and rapidly growing \acp{mbh} \cite{Barausse:2020mdt}.
On the other hand, as these two \acp{mbh} gradually approach each other and orbit around each other, they emit \ac{gw} \cite{Begelman1980}.
Therefore, the detection of \acp{gw} signals from \acp{mbhb} offers an intriguing window for direct observation of the history of \acp{mbhb}, including the inspiral, merger, and ringdown stages as two \acp{mbh} approach each other.
This, in turn, can provide insights into the \acp{mbh} growth of \acp{mbhb} and the evolution of galaxies over cosmic time \cite{Amaro-Seoane2017, eLISA:2013xep, Barausse:2014oca}.
These \ac{gw} signals can be detected by \acp{pta} \cite{1990ApJ...361..300F} and future space-borne \ac{gw} detectors.
Several \ac{pta} missions\cite{EPTA:2023fyk, Tarafdar:2022toa, NANOGrav:2023gor, Reardon:2023gzh, Xu:2023wog} have recently reported a possible detection of a \ac{sgwb}, which is most plausibly produced by an astrophysical population of merging \acp{mbhb} \cite{EPTA:2023xxk, NANOGrav:2023hfp}.
It should be noted that while space-borne \ac{gw} detectors are expected to detect individual merging \acp{mbhb},  \acp{pta} may detect a \ac{sgwb} from a population of \acp{mbhb} during the operation of space-borne \ac{gw} detectors \cite{Steinle:2023vxs}.


This paper will primarily focus on the detection of a single \ac{mbhb} using future space-borne detectors.
Several space-borne \ac{gw} detectors are proposed to operate since the mid-2030s, including TianQin, which will orbit around Earth \cite{Luo2016},
as well as LISA \cite{Danzmann:1997hm, Amaro-Seoane2017} and
Taiji \cite{Gong:2014mca, Hu:2017mde}, both orbiting around the Sun.
These detectors will consist of three spacecraft arranged in a triangular formation and will measure the \acp{gw} in the millihertz (mHz) frequency band.
Detecting \acp{mbhb} with masses ranging from $10^3$ to $10^8\, \mSun$ is one of the primary scientific objectives for these space-borne \ac{gw} detectors\cite{Klein:2015hvg, Colpi2019TheGV, Wang2019}.

Utilizing \acp{gw} opens up significant opportunities for advancing our understanding of \acp{mbhb}.
For instance, detecting \acp{mbhb} at high redshift can provide insights into their formation and evolution \acp{mbh}, as well as the co-evolution history of \acp{mbh} with their host galaxies \cite{Sesana:2010wy, Klein:2015hvg, Ricarte:2018mzn, Bonetti:2018tpf}.
Furthermore, the identification of \acp{gw} from different merger scenarios of \acp{mbhb} can help to directly address the ``final-parsec problem'' \cite{Zhu:2023imz}, which refers to the theoretically proposed stall in the dynamical evolution of \acp{mbhb} when they reach the innermost parsec of a galaxy \cite{Milosavljevic:2001vi}.
Additionally, \acp{gw} from \acp{mbhb} can serve as standard sirens for measuring the expansion history of the universe \cite{Zhu:2021aat, Zhu:2021bpp}.
While the previous scientific prospects have relied on the rough ``order of magnitude'' estimates of source parameter errors using the \ac{fim} method \cite{Wang2019, Porter2015}, it is crucial to demonstrate the ability to extract accurate astrophysical information from real observation data.
Parameter inference within the Bayesian framework is a widely adopted method for extracting such information.

\citet{Marsat2021} conducted a full Bayesian parameter estimation analysis for both \ac{mbhb} and stellar-mass black hole binary systems, considering the full frequency response of LISA \cite{Marsat2018}. They took into account the correspondence between time and frequency in their analysis.
When focusing solely on the dominant harmonic of the \ac{gw} waveform, the authors discovered two significant degeneracies.
The first is related to the sky localization of the source and its reflected position in the LISA plane.
The second concerns the distance and inclination angle of the source.
However, these degeneracies were effectively broken when higher harmonics were included in the analysis \cite{Marsat2021}.
Building upon the work of \citet{Marsat2021}, \citet{Katz2020} expanded the analysis to include various \acp{mbhb} signals with aligned spins.
They considered the presence of instrument noise in LISA observations and employed a graphics processing unit (GPU) to speed up the calculations.
It is worth noting that the introduction of noise can introduce biases in parameter recovery, with extrinsic parameters being more affected than intrinsic parameters \cite{Weaving2023}.
In addition to the aforementioned studies, \citet{Cornish2020b} developed several techniques for detecting \ac{mbhb} with LISA.
They found that the Fisher matrix method provides a poor approximation to the posterior distribution of extrinsic parameters, such as position.
Other pipelines, software, and methods have also been proposed for parameter estimation of \ac{mbhb} using single space-borne detectors, 
for example, some proposed a fully-automated and GPU-accelerated pipeline \cite{Katz:2021uax}, 
a parameter inference system named as \texttt{LitePIG} \cite{Wang:2022apn}, 
and the use of normalizing flows \cite{Du:2023plr}.

The above works primarily focus on a single space-borne detector.
However, since the multiple space-borne detectors are planned to be launched at a similar time, their mission times will overlap.
Hence, it is natural to consider the network between different detectors \cite{Gong:2021gvw, Li:2023szq}.
Some works have proven that the network of multiple detectors can significantly improve the localization and other parameters of different sources \cite{Shuman:2021ruh, Torres-Orjuela:2023hfd, Lyu:2023ctt}.
The degree of improvement in uncertainties or accuracies will vary with different parameters and depend on the source parameters.
In a study by \citet{Shuman:2021ruh}, which focused on extrinsic parameters, it was observed that the network of different space-borne detectors led to several orders of magnitude improvements in sky location and an order of magnitude improvement in luminosity distance for high-mass binaries.
However, no improvement in luminosity distance was noted for low-mass binaries.

This study aims to perform comprehensive parameter estimation of \acp{mbhb} by utilizing the combined observational capabilities of TianQin and LISA through a Bayesian analysis method.
Our investigation focuses on studying the impact of various parameters on the constraints of \ac{mbhb} parameters. 
Furthermore, we assess the constraint abilities of both detectors by analyzing two reported \ac{mbhb} systems, namely OJ287 and Tick-Tock, while also examining the parameter estimation performance in the presence of instrument noise.
Similar to the work of \cite{Shuman:2021ruh}, our waveform model for \ac{mbhb} systems is based on the \phd model, which is a complete inspiral-merger-ringdown waveform featuring aligned spins \cite{Khan2016}. 
To incorporate the effects of the motion of different detectors, we consider the full frequency response \cite{Marsat2018, Marsat2021} in both the mock data and parameter estimation process.

The remainder of the article is structured as follows.
In section \ref{sec:theory} we briefly introduce the waveform and the response function of the detector used in this study. 
Section \ref{sec:methodology} describes the statistical methods.
Next, in section \ref{sec:results}, we present the outcomes of the posterior analysis.
Finally in section \ref{sec:conc} we give our discussions and conclusions.

\section{Background} 
\label{sec:theory}

In this section, we describe the detection model that we use in the Bayesian parameter estimation. 
These include the choice of the waveform model for the \ac{gw} source, and the response and noise models for the detectors.

\subsection{The \acp{mbhb} systems}
\label{sec:mbhb}

In the study of \acp{mbhb}, waveform models play a crucial role in understanding the \ac{gw} signals emitted by these systems.
The quasicircular orbit of \acp{mbhb} can be characterized by several intrinsic parameters, which describe the dynamics of the binary system \cite{Cornish:2006ms, Shuman:2021ruh}.
The intrinsic parameters include the individual masses $m_i$ and the spin vectors $\boldsymbol{S}_i$ for the primary and secondary \ac{mbh}.
Many \ac{gw} models and some inference codes often employ the reparameterization of the intrinsic parameter to expedite the convergence speed of sampling.
This includes the use of the chirp mass and symmetric mass ratio, denoted as $\Mc=\frac{(m_1 m_2)^{3/5}}{(m_1+m_2)^{1/5}}$ and $\eta=\frac{(m_1 m_2)}{(m_1+m_2)^2}$, respectively.
It is important to note that $m_1 \geq m_2$ is usually required.
Additionally, the dimensionless spins $\boldsymbol{a}_i = \boldsymbol{S}_i / m_i^2$ are also considered.
In terms of the waveform, two additional parameters $\{t_c, \phi_c\}$ are needed to represent the coalescence time and the corresponding orbital phase of the binaries.

Apart from these intrinsic parameters, some extrinsic parameters, which are tied to the response function of the detectors \cite{Cornish:2006ms, Shuman:2021ruh}, are required to describe the waveform in the observer's frame.
If the source is not facing the observers directly ($\iota = 0$), the inclination angle $\iota$ is necessary, which denotes the angle between the orbital angular momentum of the source and the line of sight to the source.
To indicate the spatial position of the source in the observer frame, 
one needs the luminosity distance $D_L$, 
as well as the ecliptic longitude and ecliptic latitude $\{\lambda, \beta\}$ of the source.
According to the general relativity, \ac{gw} has two polarizations, namely the $+$ and the $\times$ modes.
Consequently, a polarization angle $\psi$ is required to describe the relative rotation between the polarization basis at the source frame and the observer frame.

The evolution of \ac{mbhb} systems can be divided into three stages: inspiral, merger, and ringdown.
The inspiral stage can be well approximated with the \ac{pn} expansion \cite{Blanchet:2013haa}, while the ringdown can be described by the quasi-normal modes \cite{Berti:2009kk}.
The accurate \ac{gw} waveforms for the highly dynamic merger process are obtained through the numerical calculation of the Einstein Field Equation, referred to as \ac{nr} \cite{Jani:2016wkt, Mroue:2013xna, Boyle:2019kee}.
Waveforms from different stages are concatenated and calibrated to formulate practical waveform models.
Currently, there are several rapidly computed \ac{gw} waveform models available, such as the \texttt{IMRPhenom} family \cite{Ajith:2007qp, Khan:2015jqa, Garcia-Quiros:2020qpx},
the \texttt{SEOBNR} family \cite{Buonanno:1998gg, Damour:2001tu, Bohe:2016gbl, Pan:2013rra, Ossokine:2020kjp},
the \texttt{TEOBResumS} family \cite{Nagar:2018zoe},
and the surrogate models like \texttt{NRSur} family \cite{Blackman:2015pia, Field:2013cfa}.
For our paper, we select the \phd \cite{Khan2016} model, which is widely utilized in data simulation and data challenge for space-borne \ac{gw} detectors \cite{Baghi:2022ucj, Ren:2023yec, Li:2023szq}, etc.

The \phd is a frequency-domain phenomenological waveform model that only contains the dominant quadrupole mode, i.e., the (2,2) mode. 
It can generate \ac{gw} signals for coalescence black hole binaries with non-precessing spins and only align-spins, denoted as $\boldsymbol{a}_i = \{0,0, S_i^z\} / m_i^2$ (in this paper we will use the symbols $a_i = S^z_i/m_i^2$ represent the spins components).

In the source frame, the \ac{gw} signal of the dominant (2,2) mode can be represented by the amplitude $A_{22}$ and the phase $\Psi_{22}$ as follows
\begin{equation}
    \begin{aligned}
        h_+ = & \, (1+\cos^2 \iota) A_{22} \cos \Psi_{22} (t), \\
        h_\times = & \, 2\cos\iota A_{22} \sin \Psi_{22} (t),
    \end{aligned}
    \label{eq:hphc}
\end{equation}
where $h_+, h_\times$ correspond to the two polarizations of the \ac{gw}, while $t$ represent the time.
It should be noted that $\Mc \rightarrow (1+{\mathrm z}) \Mc$ used in this equation refers to the redshifted chip mass, with ${\mathrm z}$ being the redshift of the source.
By adopting the traceless-transverse gauge, the \ac{gw} strain tensor in the observer frame can be expressed as
\begin{equation}
    \begin{aligned}
    \bfh^{\rm TT} =& \; h_+ (\bfepsilon_+ \cos 2\psi + \bfepsilon_\times \sin 2\psi) 
    \\
    &  +  h_\times (\bfepsilon_\times \cos 2\psi - \bfepsilon_+ \sin 2\psi),
    \end{aligned}
\end{equation}
where $\epsilon_{+, \times}$ are the polarization tensors, given by
\begin{equation}
    \begin{aligned}
        \bfepsilon_+ = & \bfu \otimes \bfu - \bfv \otimes \bfv, \\
        \bfepsilon_\times = & \bfu \otimes \bfv + \bfv \otimes \bfu.
    \end{aligned}
    \label{eq:pol_tensor}
\end{equation}
Here, vectors $\bfu$ and $\bfv$ along with the \ac{gw} propagation vector $\bfk$ can be described in the observer frame using spherical coordinates
\begin{align}
    \bfu = & \{\sin \lambda, -\cos\lambda, 0\}, \\
    \bfv = & \{-\sin \beta \cos\lambda, -\sin \beta \sin \lambda, \cos\beta\}, \\
    \bfk = & \{-\cos\beta \cos\lambda, -\cos\beta \sin\lambda, -\sin \beta\}.
\end{align}

Finally, in the frequency domain, the expression of the strain for the 22 mode can be written as 
\begin{equation}
    \tdbfh(f) = \bfP_{22} \tdh_{22}(f),
\end{equation}
where
\begin{equation}
    \begin{aligned}
        \bfP_{22} = & \frac{1}{4} \sqrt{\frac{5}{\pi}}  
        \bigg[
        \cos^4\frac{\iota}{2} \;
        (\bfepsilon_+ + \bfepsilon_\times ) e^{-2\ui \psi}
        \\
        &  +  \sin^4 \frac{\iota}{2} \;
        (\bfepsilon_\times - \bfepsilon_+ ) e^{2\ui \psi}
        \bigg],
    \end{aligned}    
\end{equation}
and $\tdh_{22}(f)$ is the waveform in the frequency domain, which only depends on the parameters $(\Mc, \eta, D_L, \phi_{c}, t_{c}, a_1, a_2 )$.

\subsection{The detection of \ac{gw}}

We consider two detectors in this work: TianQin \cite{Luo2016} and LISA \cite{Amaro-Seoane2017}. 
These are representative examples of the laser interferometer \ac{gw} detectors that seek to launch in the 2030's \cite{Gong:2021gvw}, with TianQin representing a geocentric mission and LISA representing a heliocentric mission.
TianQin is comprised of three identical drag-free satellites that form a nearly equilateral triangular constellation. 
Each pair of satellites is linked by two one-way infrared laser beams which can be used, together with the intra-satellite laser links, to synthesize up to three Michelson interferometers. 
TianQin adopts a circular geocentric orbit with a radius of $10^5$ km, hence the arm length of each side of the triangle is approximately $1.73 \times 10^5$ km.
The detector plane has been chosen to face RX J0806+1527, which is believed to be a Galactic white dwarf binary system with the sky position of $( \theta_{s}=- 4.7^{\circ}, \phi_{s}=120.5^{\circ} )$ in the ecliptic coordinates. \cite{Strohmayer2008}. 
The proposed orbit for LISA is an Earth-trailing heliocentric orbit between 50 and 65 million km from Earth, with a mean inter-spacecraft separation distance of 2.5 million km \cite{Amaro-Seoane2017}.
Although the orbit motion of TianQin and LISA are completely different, both missions use a nearly normal triangle constellation, and this leads to a common expression for their response to \acp{gw}.

The detection or the single link response of \ac{gw} with the laser interferometers can be represented by the laser frequency shift between the sender $s$ and the receiver $r$.
The link response can be described by the relative frequency shift \cite{estabrook1975response, armstrong_time-delay_1999, Vallisneri:2004bn}
\begin{equation}
    \begin{aligned}
        y_{sr}(t) \simeq & \; \frac{1}{2 (1 - \bfk \cdot \bfn_{sr})} 
    \bfn_{sr} \cdot
    \big[ \bfh(t - L - \bfk \cdot \bfp_s ) 
    \\
    & - \bfh(t - \bfk \cdot \bfp_r) \big]
    \cdot \bfn_{sr},
    \end{aligned}    
    \label{eq:response_t}
\end{equation}
where $L$ and $\bfn$ is the delay\footnote{For convenience, we have chosen the natural unit so the light speed is $c = 1$.} 
and the link unit vector from $s$ to $r$, $\bfp_s$, and $\bfp_r$ are the positions of the spacecraft in the observer frame.
Here, $\bfn, \bfp_s$ and $\bfp_r$ are evaluated at the same time $t$.
Considering that the waveform is calculated in the frequency domain, the response can be written as
\begin{equation}
    \tdy_{sr} = \Tsr(f) \tdh(f),
\end{equation}
where $\Tsr(f)$ is the transfer function.
Because of the evolution of the detector constellation, the transfer function is temporal- and frequency-dependent. 
Applying the perturbative formalism in Ref. \cite{Marsat2018} to leading order in the separation of timescales, we have the transfer function for the (2,2) mode
\begin{equation}
    \mathcal{T}^{22}_{sr}(f) = G^{22}_{sr}(f,t_{f}^{22}),
\end{equation}
where
\begin{equation}
    \begin{aligned}
        G^{22}_{sr}(f,t) =& \frac{\ui \pi fL}{2}{\rm sinc}\Big[ \pi f L(1- \bfk\cdot \bfn_{sr}) \Big ] \\
        & \times \exp \Big [\ui \pi f (L+ \bfk \cdot (\bfp^L_{r}+\bfp^L_{s}) ) \Big ] \\
        & \times \exp(2 \ui \pi f \bfk \cdot \bfp_0) \; \bfn_{sr} \cdot \bfP_{22} \cdot \bfn_{sr}
        \label{kernel}
    \end{aligned}
\end{equation}
is the kernel,
$\bfp_s^L$ and $\bfp_r^L$ are the positions of spacecraft $s$ and $r$ measured from the center of the constellation $\bfp_0$, 
i.e., $\bfp_{s,r} = \bfp_0 + \bfp_{s,r}^L$.
The global factor $\exp(2\ui\pi f \bfk \cdot \bfp_0)$ is the Doppler modulation in \ac{gw} phase and the $\bfn_{sr} \cdot \bfP_{22} \cdot \bfn_{sr}$ term is the projection of the \ac{gw} tensor on the interferometer axes, which is associated with the antenna pattern function.
The relationship between the evolution of the signal's frequency and time will be
\begin{equation}
    t_{f}^{22} = -\frac{1}{2\pi}\frac{{\rm d}\Psi_{22}}{{\rm d}f}.
    \label{eq:tf}
\end{equation}
In this way, one can get the full frequency response of the signal with the consideration of the motion of the detectors.

For the space-borne laser interferometers, \ac{tdi} technique is proposed to cancel the laser frequency noise \cite{Tinto1999, Tinto2005, Armstrong1999, Estabrook2000, Dhurandhar2002}. 
In the work, we only consider first-generation \ac{tdi} and adopt a rigid approximation for the constellation, where delays are all constant and equal to the armlength $L$.

We use the notation for the laser frequency shift: $y_{sr,nL} = y_{sr}(t - nL)$, which denotes an incoming \ac{gw} across the link between spacecraft $s$ and $r$.
From delayed combinations of the single-link observables $y_{sr}$, the TDI-1st generation Michelson-like channel observables $X$ \cite{Vallisneri2005} read
\begin{equation}
    \begin{aligned}
        X =&\; y_{31} + y_{13,L} + y_{21,2L} + y_{12,3L}  \\
        & - ( y_{21} + y_{12,L} + y_{31,2L} + y_{13,3L} )
    \end{aligned}
\end{equation}
and the other Michelson-like observables $Y$, $Z$ being obtained by cyclic permutation.
With the $X, Y, Z$ channel, one can construct uncorrelated combinations, i.e., the $A$, $E$, and $T$ channels \cite{Prince2002}.
In the frequency domain, they are \cite{Prince2002}
\begin{align}
    \tdA =& \frac{1}{\sqrt{2}}( \tdZ - \tdX) 
    \nn \\
    =& \frac{z^{2}-1}{\sqrt{2}} \Big[(1+z) ( \tdy_{31} + \tdy_{13} )
    \nn \\
    &- \tdy_{23} - z \tdy_{32} - \tdy_{21}- z \tdy_{12} \Big] \,,
    \label{eq:A} \\
    \tdE =& \frac{1}{\sqrt{6}} ( \tdX - 2\tdY + \tdZ ) \nn\\
    =& \frac{z^{2}-1}{\sqrt{6}} \Big[ (1-z)( \tdy_{13} - \tdy_{31} )+ (2+z) ( \tdy_{12} - \tdy_{32} ) 
    \nn\\
    &+ (1+2z) ( \tdy_{21} - \tdy_{23} ) \Big]\,,
    \label{eq:E} \\
    \tdT =& \frac{1}{\sqrt{3}} \left( \tdX+\tdY+\tdZ \right) 
    \nn\\
    =&\frac{1}{\sqrt{3}}(z^{2}-1)(1-z)
    \nn\\
    & \times ( \tdy_{21} - \tdy_{12} + \tdy_{32} - \tdy_{23} + \tdy_{13} - \tdy_{31})\,,
    \label{eq:T}
\end{align}
where $z\equiv \exp[2\ui \pi fL]$ \cite{Marsat2021}.

Apart from the response, one also needs to specify the noise model for the detectors.
For TianQin and LISA, the noise \ac{psd} in the \ac{tdi} channels ($A, E, T$) are
\bea S_n^A&=&S_n^E=8 \sin^2{\omega L}\Big[4(1+\cos{\omega L}+\cos^2{\omega L})S_{\rm acc}\nn\\
&&\qquad+ (2+\cos{\omega L})S_{\rm pos}\Big]\,,\nn\\
S_n^T &=&32\sin^2{\omega L}\sin^2{\frac{\omega L}{2}}\Big[4\sin^2{\frac{\omega L}{2}}S_{\rm acc}+S_{\rm pos}\Big]\,,\label{eq:TQ_PSD}\eea
where $\omega = 2\pi f$, 
$S_{\rm acc}$ stands for the accelerated noise,
$S_{\rm pos}$ is the position noise.
The armlength for TianQin and LISA are $L_{TQ} \approx 1.73 \times 10^5$ km \cite{Luo2016} and $L_{LISA} = 2.5 \times 10^6$ km \cite{Babak:2021mhe}, respectively.
In the unit of relative frequency, 
the noise parameters for TianQin are
\begin{align}
    \sqrt{S_{\rm acc}} (f) =& \frac{\sqrt{S_a}}{2\pi f c}
    \sqrt{1+\frac{0.1 \rm mHz}{f}}, 
    \\
    \sqrt{S_{\rm pos}} =& \sqrt{S_x} \frac{2\pi f}{c},
\end{align}
where $\sqrt{S_a} = 10^{-15} \rm m/s^2/Hz^{1/2}$ and $\sqrt{S_x} = 10^{-12} \rm m/Hz^{1/2}$ \cite{Luo2016}.
Similarly, for LISA, the noise parameters are
\begin{align}
    \sqrt{S_{\rm acc}} (f) =& \frac{\sqrt{S_a}}{2\pi f c}
    \sqrt{1+\left( \frac{0.4 \rm mHz}{f} \right)^2} \sqrt{1+\left(\frac{f}{8 \rm mHz}\right)} . \\
    \sqrt{S_{\rm pos}} = & \sqrt{S_x}  \frac{2\pi f}{c} \sqrt{1 + \left( \frac{2 \rm mHz}{f} \right)^4}, 
\end{align}
where $\sqrt{S_a} = 3\times 10^{-15} \rm m/s^2/Hz^{1/2}$ and $\sqrt{S_x} = 15 \times 10^{-12} \rm m/Hz^{1/2}$ \cite{Babak:2021mhe}.

For LISA, one also needs to add the Galactic foreground noise \cite{Babak:2021mhe}:
\bea S_{\rm Gal} = A f^{-7/3} e^{-(\frac{f}{f_1}) ^{\alpha} }
\frac1{2} \Big[1.0 + \tanh(-\frac{f-f_k}{f_2}) \Big]\,,  \label{eq:SGal}\eea
where $A = 1.14\times 10^{-44}, \, \alpha=1.8, \,  f_2 = 0.31\, \rm{mHz}\,$, and $f_1$ and $f_k$ depend on the total observation time as more signals can be resolved with longer observation time.


\section{The statistical method}
\label{sec:methodology}

The parameter estimation in this work is done with the Bayesian inference, the parameter space is sampled with a type of \ac{mcmc} method.

\subsection{Bayesian inference} \label{sec:bayes theorem}

Our analysis is based on Bayes' theorem, with which the posterior probability that the observed data $\data$ and the background hypothesis $H$ favor a model characterized by a particular set of parameter values $\params$ is given by
\bea p(\params|\data,H) = \frac{p(\data|\params,H) \, p(\params|H)}{p(\data|H)}\,,\label{eq:bayestheo}\eea
where $p(\params|\data,H)$ is the posterior, $p(\data|\params,H)$ is the likelihood, $p(\params|H)$ is the prior and $p(\data|H)$ is the evidence. Since we will fix the noise and \ac{gw} signal models in this study, the evidence can be seen as a normalization constant that does not need an explicit calculation.

In general, the \ac{gw} data $\data$ is a superposition of noise $\noise$ and a possible signal $h$. Assuming that the noise is Gaussian and stationary, the logarithm of likelihood can be written as:
\bea \ln \mathcal{L}  =  \left\langle -\frac{1}{2} \left(\data-h(\params)|\data-h(\params)\right) \right\rangle + {\rm const.}\,,\label{eq:ll}\eea
where $h$ is the signal model, which contains the waveform model and the detector's response to the signal. 
The noise-weighted inner product is defined as
\bea(a|b) = 2\int_0^\infty\frac{\tilde{a}(f)\tilde{b}(f)^* + \tilde{a}(f)^*\tilde{b}(f)}{S_n(f)} {\rm d}f\,,\label{eq:innerprod}\eea
where $S_n(f)$ is the one-sided \ac{psd} of the noise. The joint likelihood function of a network of detectors can be written as the product of the likelihood function for individual detectors. So the logarithm of the joint likelihood is given by:
\bea\ln\mathcal{L}_{\rm joint} = \ln\mathcal{L}_{\rm TianQin} + \ln\mathcal{L}_{\rm LISA} \,.\label{eq:lljoint}\eea
The \ac{snr} is defined as:
\bea{\rm SNR} = \sqrt{(h|h)}\,,\label{eq:SNR}\eea
And the joint \ac{snr} of the TianQin-LISA network is defined as:
\bea{\rm SNR}_{\rm joint}=\sqrt{ {\rm SNR}_{\rm TianQin}^2+{\rm SNR}_{\rm LISA}^2}\,.\label{eq:total-SNR}\eea
For our analysis, the prior knowledge is limited.
The priors for the eleven parameters are chosen as flat in the ranges specified in Table \ref{tb:priorinfo}.
We will see that since the parameters are well constrained in a small range of parameter space, the choice of prior distribution has little impact on the posterior.

\begin{table}[htbp]
    \centering
    \caption{The priors for the eleven parameters are chosen as flat in the following ranges.}
    \label{tb:priorinfo}
    \renewcommand{\arraystretch}{1.45}
        \begin{tabular}{|c|c|c|}
            \hline
            Parameter & Lower Bound & Upper Bound \\
            \hline
            $\Mc^*$ $[\mSun]$ & $10^3 / 10^5$ & $10^7 / 10^9$ \\
            \hline
            $\eta$ & 0.01 & 0.25 \\
            \hline
            $D_L$ $[{\rm Mpc}]$ & $0.01$ & $1\times10^5$ \\
            \hline
            $\phi_\text{ref}$ $[{\rm rad}]$ & 0.0 & $2\pi$ \\
            \hline
            $t_\text{ref}$ $[{\rm s}]$& $0.0$ & $3.15576\times10^8$ \\
            \hline
            $\lambda$$[{\rm rad}]$ & 0.0 & $2\pi$ \\
            \hline
            $\beta$$[{\rm rad}]$ & 0.0 & $\pi$ \\
            \hline
            $\psi$$[{\rm rad}]$ & 0.0 & $\pi$ \\
            \hline
            $\iota$$[{\rm rad}]$ & 0.0 & $\pi$ \\
            \hline
            $a_1(a_2)$ & -0.99 & 0.99 \\
            \hline
        \end{tabular}
        \\
        {\small $^*$Note that the masses of OJ287 and Tick-Tock are much more massive than the six examples, then the range of priors of the two systems' chirp masses are larger.}
\end{table}

\subsection{MCMC algorithm}
\label{sec:MCMC}

As discussed above, the form of the likelihood function remains indeterminate analytically, and the parameter space is up to 11 dimensions. 
In general, \ac{mcmc} methods are adept at exploring high-dimensional parameter spaces with minimal computational expenses \cite{MacKay2003}. 
In this paper, we employ an \ac{mcmc} sampler based on the \texttt{emcee}\footnote{\url{https://emcee.readthedocs.io/en/stable/index.html}} packages, which provide a \texttt{Python} implementation of an affine-invariant \ac{mcmc} ensemble sampler \cite{Goodman2010}.

\section{Results}
\label{sec:results}

In this section, we present a comprehensive description of the exemplar \ac{gw} sources employed in our computation and thoroughly examine the outcomes of the parameter estimation process.

\subsection{Example sources}

To facilitate the investigation conducted in this study, we introduce a fiducial source, denoted as F0, along with five distinct variations labeled FP (facing pole), AF (almost face-on), AS (anti-spin), HM (higher mass-ratio), and LM (light mass). 
These variants correspond to sources positioned at the pole direction of TianQin (further details provided below) and exhibit alterations in inclination angle, spin orientation, mass hierarchy, and mass magnitude, respectively. 
The parameter values for the F0 source are selected to ensure high \ac{snr} on both TianQin and LISA detectors.
Conversely, the remaining variants undergo modifications in terms of sky direction, inclination angle, spin, mass ratio, and masses. 
For instance, the sky direction of FP is aligned with J0806, thereby enabling optimal response from TianQin. 
Key parameters pertaining to these sources can be found in Table \ref{params1}, while those not listed are assigned random values that remain consistent across all five sources. 
These include the coalescence phase ($\phi_c=1.32$ rad), polarization angle ($\psi=1.23$ rad), and redshift ($z=2$). 
The luminosity distance is derived from the redshift within the base-$\Lambda$CDM model, using the cosmological parameters $H_0=67.4 \rm~km/s/Mpc$ and $\Omega_m=1-\Omega_{\Lambda}=0.315$ from the Planck 2018 result \cite{Planck:2018vyg}. 

In addition to the fiducial sources, we also examine two \ac{mbhb} candidates from \ac{em} observation, namely OJ287 \cite{Sillanpaa1988,Lehto1996,Valtonen2008} and Tick-Tock \cite{Oh2015, Jiang:2022aek}. 
Part of the parameters associated with the two sources is presented in Table \ref{params2}, with the remaining parameters assumed to match those of the F0 case.

\begin{table}[t]
    \centering
    \caption{Parameters of the fiducial source and its variants.}
    \label{params1}
    \begin{tabular}{|c|c|c|c|c|c|c|}
        \hline
        Abbreviation        & F0          & FP        & AF              & AS            & HM              & LM       \\
        \hline
        $m_1$ ($10^{6}\mSun$)       & $1.50$    & $1.50$   & $1.50$  & $1.50$        & \ccr{$10.0$}        & \ccr{$0.15$} \\
        $m_2$ ($10^{6}\mSun$)       & $1.00$    & $1.00$   & $1.00$  & $1.00$        & $1.00$        & \ccr{$0.10$} \\
        $a_1$                       & $0.96$    & $0.96$   & $0.96$  & \ccr{$-0.96$}       & $0.96$        & $0.96$ \\
        $a_2$                       & $0.94$    & $0.94$   & $0.94$  & \ccr{$-0.94$}       & $0.94$        & $0.94$ \\
        $\lambda$ (rad)             & $1.43$    & \ccr{$2.10$}   & $1.43$  & $1.43$        & $1.43$        & $1.43$ \\
        $\beta$ (rad)               & $0.52$    & \ccr{$-0.08$}  & $0.52$  & $0.52$        & $0.52$        & $0.52$ \\
        $\iota$ (rad)               & $1.46$    & $1.46$       & \ccr{$0.46$}  & $1.46$        & $1.46$        & $1.46$ \\
        \hline
        ${\rm SNR}_{\rm TQ}$        & 1340      & 2071     & 3398    & 532           & 398           & 194 \\
        ${\rm SNR}_{\rm LISA}$      & 1560      & 2068     & 3546    & 701           & 371           & 192 \\
        \hline
    \end{tabular}
\end{table}

\begin{table}[htbp]
    \centering
    \caption{Parameters of two MBHB candidates, i.e., OJ287 and Tick-Tock.} 
    \label{params2}
    \begin{tabular}{|c|c|c|}
        \hline
        Source & OJ287 & Tick-Tock \\
        \hline
        $m_1$ ($10^{8}\mSun$)       & \ccr{$1.30$}    & \ccr{$1.60$} \\
        $m_2$ ($10^{8}\mSun$)       & \ccr{$1.20$}    & \ccr{$0.40$} \\
        $D_L$ (Gpc)                 & \ccr{$1.62$}    & \ccr{$0.38$} \\
        $\lambda$ (rad)             & \ccr{$0.73$}    & \ccr{$1.15$} \\
        $\beta$ (rad)               & \ccr{$0.01$}    & \ccr{$0.20$} \\
        \hline
        ${\rm SNR}_{\rm TQ}$        & 991       & 2879 \\
        ${\rm SNR}_{\rm LISA}$      & 2138      & 9166 \\
        \hline
    \end{tabular}
\end{table}

\begin{figure}
    \centering
    \includegraphics[width=.99\linewidth]{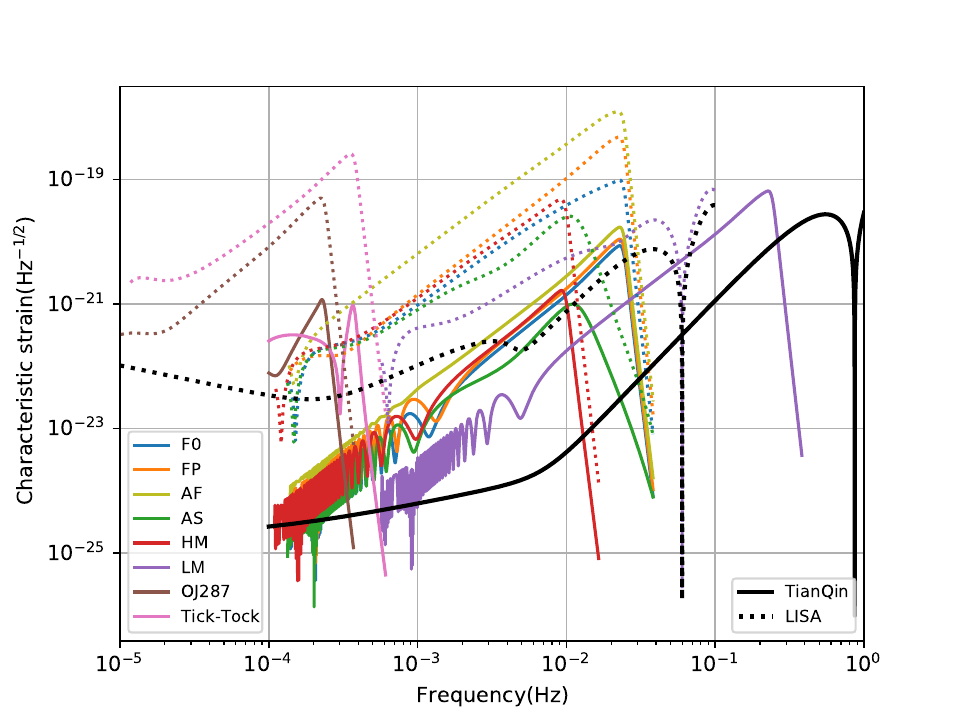}
    \caption{Characteristic strain for the example sources and the characteristic noise \acp{psd} for the detectors. 
    The solid lines are for TianQin and the dashed lines are for LISA. 
    The black lines are for the characteristic noise \ac{psd} and the dashed lines are for the characteristic strain from different sources in the \ac{tdi} channel $A$.
    More details about the sources can be found in the text.}
    \label{fig:signal}
\end{figure}

The plot in Fig. \ref{fig:signal} displays the characteristic strains of the example sources along with the characteristic noise power spectral densities (PSDs) for the detectors. These PSDs are defined as follows:
\begin{equation}
    \begin{aligned}
        \tilde{h}_{A,E,T}^{c}(f) =& 2 f \tilde{h}_{A,E,T}(f), \\
        S^{A,E,T}_{c}(f) =& f S^{A,E,T}_{n}(f).
    \end{aligned}    
    \label{eq:defhc}
\end{equation}
Given that the $A$ and $E$ channels exhibit similar characteristics, mostly noise at low frequencies, only the $A$ channel is depicted in Fig. \ref{fig:signal}. 
With this definition, the region between the characteristic strains and the characteristic noise represents the \ac{snr} of the corresponding signal at the detectors. 
The sensitivity bands are assumed to be [$10^{-4}$, 1] Hz for TianQin \cite{Mei2021} and [$10^{-5}$, 0.1 ] Hz for LISA \cite{Katz2022}. 
Note that there is strong modulation on the low-frequency signals detected by TianQin, due to the 3.65-day orbit motion of TianQin spacecraft around the center of Earth.

In this study, we exclusively utilize the final three months of data for each source and assume that TianQin can detect all of this data. 
This assumption implies that all sources merge precisely during TianQin's three-month detection window. 
Although this assumption may not be entirely realistic, it serves our purpose of testing the influence of parameter values on parameter estimation. 
By adopting this approach, the initial frequency for F0 case starts at approximately $6.84\times10^{-5}$ Hz and merges at around $1.62\times10^{-2}$ Hz. 
One can expect that TianQin's \ac{snr} will be smaller than that of LISA.

In the case of the FP scenario, the situation is similar, but both TianQin and LISA exhibit very close \acp{snr} due to TianQin's optimal response to the source located in the pole. 
Comparing the AF case with the F0 case, we observe that the \acp{snr} of the AF case are more than twice as high as that of the F0 case.
This discrepancy primarily arises from the choice of the parameter value $\iota$. 
According to Eq.~\eqref{eq:hphc}, as $\iota \rightarrow \pi/2$ (edge-on configuration), $\cos\iota \rightarrow 0$, causing the $\times$ mode to vanish and the $+$ mode to become weaker.

Regarding the AS case, both TianQin and LISA exhibit \acp{snr} that are roughly less than half of those in the F0 case, thanks to the lower merger frequency caused by the anti-aligned spin. 
In the HM case, the merging of \ac{mbhb} occurs precisely where the Galactic confusion noise is most significant for LISA, resulting in a slightly higher \ac{snr} for TianQin. 
Furthermore, LM merges at approximately 0.16 Hz, giving TianQin a slightly higher \ac{snr} than LISA, as the latter is limited to frequencies below 0.1 Hz. 
Additionally, there is a substantial disparity in the \acp{snr} of OJ287 and Tick-Tock between TianQin and LISA. 
This difference primarily arises from the varying frequency bands of TianQin and LISA. 
LISA can acquire more information about the inspiral phase of the two signals at lower frequencies.

\begin{table*}
    \centering
    \caption{The constraint results of different parameters for the different \ac{mbhb} systems measured with different detectors, i.e, TianQin/LISA/Joint (the network of TianQin and LISA).
    The errors are within 1$\sigma$ C.L.}
    \label{tb:error_emcee}
    \renewcommand{\arraystretch}{1.45}
    \setlength{\tabcolsep}{1pt}
    \begin{tabular}{@{}|c|c|c|c|c|c|c|c|c|c|c|@{}}
        \hline
        Parameter & Detector &F0 & FP & AF & AS & HM & LM & OJ287 & Tick-Tock & F0+noise\\
        \hline
        \multirow{3}{*}{\parbox[t]{0.08\linewidth}{$\frac{\Delta\Mc}{\Mc}$ $(\times 10^{-4})$}}
        & TQ & $-0.63_{-1.68}^{+1.75}$ & $ 0.05_{-1.05}^{+1.24}$ & $ 0.0_{-0.77}^{+0.70}$ & $-0.24_{-4.32}^{+4.84}$ & $-0.65_{-6.61}^{+6.02}$ & $-0.01_{-0.44}^{+0.44}$ & $-29.6_{-18.7}^{+13.8}$ & $11.5_{-44.9}^{+55.0}$ & $-0.17_{-1.71}^{+2.37}$ \\
        & LISA & $-0.07_{-1.12}^{+1.08}$ & $-0.06_{-0.92}^{+0.79}$ & $ 0.02_{-0.52}^{+0.48}$ & $-0.07_{-2.97}^{+2.42}$ & $-0.44_{-2.83}^{+3.51}$ & $-0.03_{-0.37}^{+0.36}$ & $-6.60_{-6.70}^{+5.10}$ & $-11.6_{-1.40}^{+1.50}$ & $ 0.57_{-1.0}^{+1.03}$ \\
        & Joint & $-0.14_{-0.86}^{+0.87}$ & $ 0.04_{-0.68}^{+0.65}$ & $ 0.0_{-0.36}^{+0.35}$ & $-0.08_{-2.39}^{+2.45}$ & $ 0.02_{-2.96}^{+2.85}$ & $-0.02_{-0.24}^{+0.25}$ & $-6.60_{-7.30}^{+4.10}$ & $-11.6_{-1.10}^{+1.10}$ & $ 0.18_{-0.92}^{+0.93}$ \\
        \hline
        \multirow{3}{*}{\parbox[t]{0.08\linewidth}{$\Delta\eta$ $(\times 10^{-4})$}}
        & TQ & $ 0.0_{-0.70}^{+0.80}$ & $ 0.0_{-1.30}^{+1.90}$ & $ 0.0_{-0.40}^{+0.40}$ & $ 0.0_{-1.80}^{+1.90}$ & $ 3.55_{-2.20}^{+2.40}$ & $ 0.0_{-0.20}^{+0.20}$ & $-16.0_{-14.1}^{+10.9}$ & $ 0.0_{-24.3}^{+29.7}$ & $ 0.0_{-0.70}^{+0.90}$ \\
        & LISA & $ 0.0_{-0.60}^{+0.60}$ & $ 0.0_{-1.70}^{+1.80}$ & $ 0.0_{-0.40}^{+0.40}$ & $ 0.0_{-2.10}^{+2.10}$ & $ 3.55_{-2.40}^{+2.40}$ & $ 0.0_{-30.1}^{+32.9}$ & $-6.0_{-4.80}^{+3.90}$ & $ 0.0_{-0.80}^{+0.80}$ & $ 0.0_{-0.50}^{+0.40}$ \\
        & Joint & $ 0.0_{-0.40}^{+0.50}$ & $ 0.0_{-1.30}^{+1.20}$ & $ 0.0_{-0.30}^{+0.30}$ & $ 0.0_{-1.0}^{+1.0}$ & $ 3.55_{-1.60}^{+1.60}$ & $ 0.0_{-0.20}^{+0.20}$ & $ 4.0_{-5.30}^{+3.0}$ & $ 0.0_{-0.60}^{+0.60}$ & $ 0.0_{-0.40}^{+0.40}$ \\
        \hline
        \multirow{3}{*}{\parbox[t]{0.08\linewidth}{$\frac{\Delta D_L}{D_L}$ $(\times 10^{-2})$}}
        & TQ & $ 0.31_{-1.72}^{+2.05}$ & $-0.02_{-0.07}^{+0.06}$ & $ 2.38_{-2.66}^{+4.37}$ & $-1.41_{-5.01}^{+4.34}$ & $ 2.99_{-10.3}^{+10.7}$ & $-0.33_{-3.29}^{+2.22}$ & $ 0.33_{-5.33}^{+2.83}$ & $ 5.17_{-6.59}^{+3.23}$ & $-0.82_{-2.82}^{+3.03}$ \\
        & LISA & $ 0.42_{-1.53}^{+1.62}$ & $-0.26_{-1.15}^{+0.92}$ & $ 3.14_{-4.60}^{+2.89}$ & $ 0.75_{-3.71}^{+2.50}$ & $-0.93_{-7.42}^{+6.81}$ & $ 0.06_{-2.83}^{+2.73}$ & $ 3.13_{-7.30}^{+7.81}$ & $ 4.31_{-4.41}^{+4.97}$ & $ 0.77_{-1.73}^{+2.14}$ \\
        & Joint & $ 0.0_{-0.19}^{+0.20}$ & $ 0.0_{-0.05}^{+0.05}$ & $ 0.03_{-0.35}^{+0.38}$ & $ 0.0_{-0.47}^{+0.49}$ & $-0.01_{-0.88}^{+0.87}$ & $ 0.05_{-1.20}^{+1.14}$ & $ 1.85_{-0.24}^{+0.21}$ & $ 0.96_{-0.06}^{+0.07}$ & $-0.30_{-0.19}^{+0.20}$ \\
        \hline
        \multirow{3}{*}{\parbox[t]{0.08\linewidth}{$\Delta\phi_{c}$ $(\times 10^{-1})$ (rad)}}
        & TQ & $-0.13_{-0.75}^{+0.69}$ & $-0.12_{-1.09}^{+1.15}$ & $ 0.20_{-4.98}^{+7.41}$ & $ 0.23_{-4.17}^{+3.73}$ & $-1.36_{-7.53}^{+9.43}$ & $-0.32_{-3.95}^{+3.81}$ & $-0.37_{-2.14}^{+2.31}$ & $ 0.39_{-5.18}^{+4.15}$ & $-0.06_{-0.72}^{+0.82}$ \\
        & LISA & $-0.16_{-0.54}^{+0.62}$ & $-0.14_{-1.16}^{+1.21}$ & $ 2.63_{-6.66}^{+5.48}$ & $-0.75_{-3.77}^{+4.52}$ & $-0.04_{-7.90}^{+8.63}$ & $ 0.01_{-4.07}^{+3.63}$ & $ 0.0_{-1.31}^{+0.87}$ & $-0.11_{-0.55}^{+0.50}$ & $-0.62_{-0.57}^{+0.57}$ \\
        & Joint & $-0.06_{-0.44}^{+0.46}$ & $ 0.07_{-0.92}^{+0.85}$ & $-0.01_{-0.30}^{+0.29}$ & $-0.07_{-3.51}^{+3.72}$ & $ 0.05_{-5.77}^{+5.86}$ & $ 0.08_{-3.62}^{+3.47}$ & $ 0.07_{-0.69}^{+0.56}$ & $-0.04_{-0.29}^{+0.28}$ & $-0.28_{-0.44}^{+0.44}$ \\
        \hline
        \multirow{3}{*}{\parbox[t]{0.08\linewidth}{$\Delta t_{c}$ (s) }}
        & TQ & $ 1.64_{-7.32}^{+6.14}$ & $ 0.87_{-5.34}^{+6.41}$ & $ 0.92_{-1.87}^{+1.75}$ & $-5.0_{-15.4}^{+11.2}$ & $ 3.35_{-83.5}^{+61.5}$ & $-3.45_{-7.48}^{+8.20}$ & $90_{-474}^{+427}$ & $-24.6_{-665}^{+791}$ & $-5.46_{-8.32}^{+11.0}$ \\
        & LISA & $-0.69_{-4.45}^{+4.40}$ & $-0.79_{-5.62}^{+6.67}$ & $-0.69_{-0.83}^{+0.82}$ & $ 1.74_{-14.4}^{+11.6}$ & $ 1.46_{-83.2}^{+63.9}$ & $ 0.19_{-7.92}^{+6.52}$ & $ 6.55_{-169}^{+254}$ & $ 2.03_{-84.8}^{+88.2}$ & $ 2.03_{-6.34}^{+5.11}$ \\
        & Joint & $ 0.12_{-1.07}^{+1.06}$ & $-0.12_{-1.66}^{+1.78}$ & $ 0.03_{-0.47}^{+0.48}$ & $ 0.19_{-7.40}^{+6.95}$ & $-0.15_{-50.8}^{+50.2}$ & $ 0.14_{-3.87}^{+3.46}$ & $-13.5_{-105}^{+129}$ & $ 5.95_{-42.8}^{+45.0}$ & $ 0.0_{-1.09}^{+1.04}$ \\
        \hline
        \multirow{3}{*}{\parbox[t]{0.08\linewidth}{$\Delta\lambda$ $(\times 10^{-2})$ (rad)}}
        & TQ & $ 0.50_{-2.32}^{+1.94}$ & $-0.60_{-2.38}^{+2.23}$ & $ 1.30_{-0.91}^{+0.78}$ & $-0.60_{-4.74}^{+5.38}$ & $ 5.30_{-11.2}^{+10.5}$ & $ 0.30_{-3.44}^{+2.53}$ & $149.6_{-16.4}^{+14.4}$ & $244.1_{-1.93}^{+1.20}$ & $ 0.30_{-2.82}^{+2.59}$ \\
        & LISA & $-1.50_{-2.50}^{+2.43}$ & $ 0.30_{-1.84}^{+1.83}$ & $-0.50_{-0.29}^{+0.25}$ & $-0.70_{-5.33}^{+4.80}$ & $ 2.90_{-12.9}^{+11.2}$ & $-0.50_{-3.27}^{+4.0}$ & $151.5_{-12.2}^{+8.70}$ & $250.5_{-7.32}^{+8.18}$ & $-1.30_{-2.70}^{+2.82}$ \\
        & Joint & $ 0.0_{-0.11}^{+0.11}$ & $ 0.0_{-0.09}^{+0.08}$ & $ 0.0_{-0.04}^{+0.05}$ & $ 0.0_{-0.25}^{+0.25}$ & $ 0.0_{-0.42}^{+0.41}$ & $ 0.0_{-0.59}^{+0.62}$ & $154.7_{-0.63}^{+0.78}$ & $244.9_{-0.07}^{+0.07}$ & $ 0.0_{-0.11}^{+0.11}$ \\
        \hline
        \multirow{3}{*}{\parbox[t]{0.08\linewidth}{$\Delta \beta$ $(\times 10^{-2})$ (rad)}}
        & TQ & $-0.60_{-2.48}^{+2.99}$ & $-0.20_{-1.93}^{+1.50}$ & $ 0.10_{-0.80}^{+0.63}$ & $ 1.80_{-5.0}^{+5.57}$ & $ 1.50_{-11.0}^{+13.9}$ & $ 1.10_{-3.26}^{+3.55}$ & $ 8.60_{-16.1}^{+16.8}$ & $41.9_{-1.47}^{+1.59}$ & $ 2.20_{-4.69}^{+3.13}$ \\
        & LISA & $ 0.0_{-1.77}^{+1.51}$ & $ 0.40_{-2.02}^{+1.72}$ & $ 0.20_{-0.23}^{+0.15}$ & $-0.30_{-3.42}^{+3.56}$ & $-0.20_{-8.68}^{+8.21}$ & $-0.10_{-2.68}^{+2.81}$ & $ 3.70_{-4.24}^{+7.25}$ & $44.0_{-3.20}^{+3.57}$ & $-0.40_{-2.0}^{+2.11}$ \\
        & Joint & $ 0.0_{-0.22}^{+0.23}$ & $ 0.0_{-0.09}^{+0.09}$ & $ 0.0_{-0.03}^{+0.03}$ & $ 0.0_{-0.56}^{+0.55}$ & $ 0.0_{-0.98}^{+0.99}$ & $-0.10_{-1.30}^{+1.40}$ & $ 3.60_{-0.42}^{+0.36}$ & $42.2_{-0.07}^{+0.07}$ & $ 0.30_{-0.23}^{+0.23}$ \\
        \hline
        \multirow{3}{*}{\parbox[t]{0.08\linewidth}{$\Delta \psi$ $(\times 10^{-3})$ (rad)}}
        & TQ & $ 1.0_{-6.76}^{+7.08}$ & $ 0.0_{-1.84}^{+1.85}$ & $21.0_{-493.5}^{+751.4}$ & $-4.0_{-18.9}^{+13.6}$ & $10.0_{-40.6}^{+32.1}$ & $-2.0_{-8.98}^{+8.74}$ & $ 2.0_{-11.2}^{+17.4}$ & $23.0_{-48.0}^{+58.7}$ & $-4.0_{-10.7}^{+9.65}$ \\
        & LISA & $ 3.0_{-6.86}^{+7.32}$ & $ 0.0_{-9.18}^{+8.81}$ & $261.0_{-662.5}^{+559.4}$ & $ 4.0_{-14.6}^{+12.7}$ & $ 0.0_{-28.0}^{+21.8}$ & $ 1.0_{-12.9}^{+12.6}$ & $15.0_{-37.1}^{+29.2}$ & $32.0_{-44.98}^{+43.4}$ & $ 4.0_{-7.30}^{+9.05}$ \\
        & Joint & $ 0.0_{-0.76}^{+0.75}$ & $ 0.0_{-0.31}^{+0.30}$ & $ 1.0_{-17.9}^{+17.5}$ & $ 0.0_{-1.78}^{+1.72}$ & $ 0.0_{-3.08}^{+3.08}$ & $ 0.0_{-4.82}^{+4.47}$ & $ 0.0_{-3.26}^{+2.73}$ & $ 0.0_{-0.85}^{+0.83}$ & $-1.0_{-0.75}^{+0.73}$ \\
        \hline
        \multirow{3}{*}{\parbox[t]{0.08\linewidth}{$\Delta \iota$ $(\times 10^{-3})$ (rad)}}
        & TQ & $ 0.0_{-0.92}^{+1.32}$ & $ 0.0_{-0.27}^{+0.31}$ & $-37.0_{-109.4}^{+68.8}$ & $ 0.0_{-2.86}^{+3.40}$ & $-3.0_{-4.59}^{+6.14}$ & $ 0.0_{-2.90}^{+3.09}$ & $ 0.0_{-0.63}^{+0.66}$ & $33.0_{-97.4}^{+36.5}$ & $-1.0_{-1.66}^{+1.68}$ \\
        & LISA & $-1.0_{-4.16}^{+3.37}$ & $ 0.0_{-0.59}^{+0.65}$ & $-80.0_{-82.4}^{+106.1}$ & $ 0.0_{-6.81}^{+6.64}$ & $ 3.0_{-17.8}^{+12.3}$ & $ 1.0_{-7.09}^{+5.95}$ & $ 2.0_{-5.91}^{+3.67}$ & $ 3.0_{-3.98}^{+3.06}$ & $-1.0_{-4.56}^{+3.82}$ \\
        & Joint & $ 0.0_{-0.33}^{+0.31}$ & $ 0.0_{-0.17}^{+0.19}$ & $-1.0_{-8.49}^{+7.72}$ & $ 0.0_{-0.73}^{+0.74}$ & $ 0.0_{-1.06}^{+1.08}$ & $ 0.0_{-2.07}^{+2.20}$ & $ 0.0_{-0.46}^{+0.43}$ & $ 0.0_{-0.07}^{+0.07}$ & $ 0.0_{-0.30}^{+0.31}$ \\
        \hline
        \multirow{3}{*}{\parbox[t]{0.08\linewidth}{$\Delta a_{1}$ $(\times 10^{-3})$}}
        & TQ & $ 0.0_{-2.28}^{+2.40}$ & $ 0.0_{-3.94}^{+3.65}$ & $ 0.0_{-1.01}^{+0.99}$ & $-1.0_{-18.3}^{+19.4}$ & $ 0.0_{-1.34}^{+1.13}$ & $ 2.0_{-18.4}^{+18.8}$ & $ 1.0_{-27.6}^{+19.3}$ & $ 0.0_{-1.33}^{+1.43}$ & $ 0.0_{-2.55}^{+2.41}$ \\
        & LISA & $ 1.0_{-2.01}^{+1.76}$ & $ 0.0_{-4.11}^{+3.92}$ & $ 0.0_{-1.15}^{+1.06}$ & $ 4.0_{-23.8}^{+20.4}$ & $ 0.0_{-1.46}^{+1.30}$ & $ 0.0_{-18.0}^{+20.3}$ & $ 1.0_{-21.8}^{+17.4}$ & $ 0.0_{-0.17}^{+0.18}$ & $ 2.0_{-1.86}^{+1.92}$ \\
        & Joint & $ 0.0_{-1.51}^{+1.43}$ & $ 0.0_{-2.92}^{+3.12}$ & $ 0.0_{-0.82}^{+0.81}$ & $ 0.0_{-17.7}^{+16.9}$ & $ 0.0_{-0.97}^{+0.94}$ & $ 0.0_{-16.8}^{+17.4}$ & $-1.0_{-14.9}^{+15.9}$ & $ 0.0_{-0.11}^{+0.10}$ & $ 1.0_{-1.43}^{+1.44}$ \\
        \hline
        \multirow{3}{*}{\parbox[t]{0.08\linewidth}{$\Delta a_{2}$ $(\times 10^{-3})$}}
        & TQ & $-1.0_{-4.96}^{+5.32}$ & $-1.0_{-6.90}^{+7.36}$ & $ 0.0_{-2.08}^{+2.04}$ & $ 1.0_{-34.8}^{+32.5}$ & $-5.0_{-23.9}^{+31.0}$ & $-3.0_{-31.3}^{+32.3}$ & $-4.0_{-25.7}^{+37.5}$ & $ 0.0_{-20.2}^{+16.3}$ & $ 0.0_{-5.15}^{+5.81}$ \\
        & LISA & $-1.0_{-3.70}^{+4.42}$ & $-1.0_{-7.09}^{+7.77}$ & $ 0.0_{-2.43}^{+2.31}$ & $-6.0_{-30.6}^{+35.8}$ & $ 0.0_{-25.0}^{+27.6}$ & $ 3.0_{-39.6}^{+27.9}$ & $-1.0_{-22.6}^{+24.9}$ & $-1.0_{-2.05}^{+1.98}$ & $-3.0_{-4.02}^{+3.87}$ \\
        & Joint & $ 0.0_{-3.08}^{+3.07}$ & $ 0.0_{-5.71}^{+5.45}$ & $ 0.0_{-1.59}^{+1.67}$ & $-1.0_{-29.2}^{+31.3}$ & $ 0.0_{-18.2}^{+18.1}$ & $ 0.0_{-29.8}^{+29.0}$ & $ 0.0_{-17.06}^{+17.7}$ & $ 0.0_{-1.11}^{+1.09}$ & $-1.0_{-3.10}^{+3.17}$ \\
        \hline
    \end{tabular}
\end{table*}

\begin{figure*}
    \centering
    \includegraphics[width=\linewidth, trim=2.8cm 2.8cm 2.4cm 2.4cm, clip]{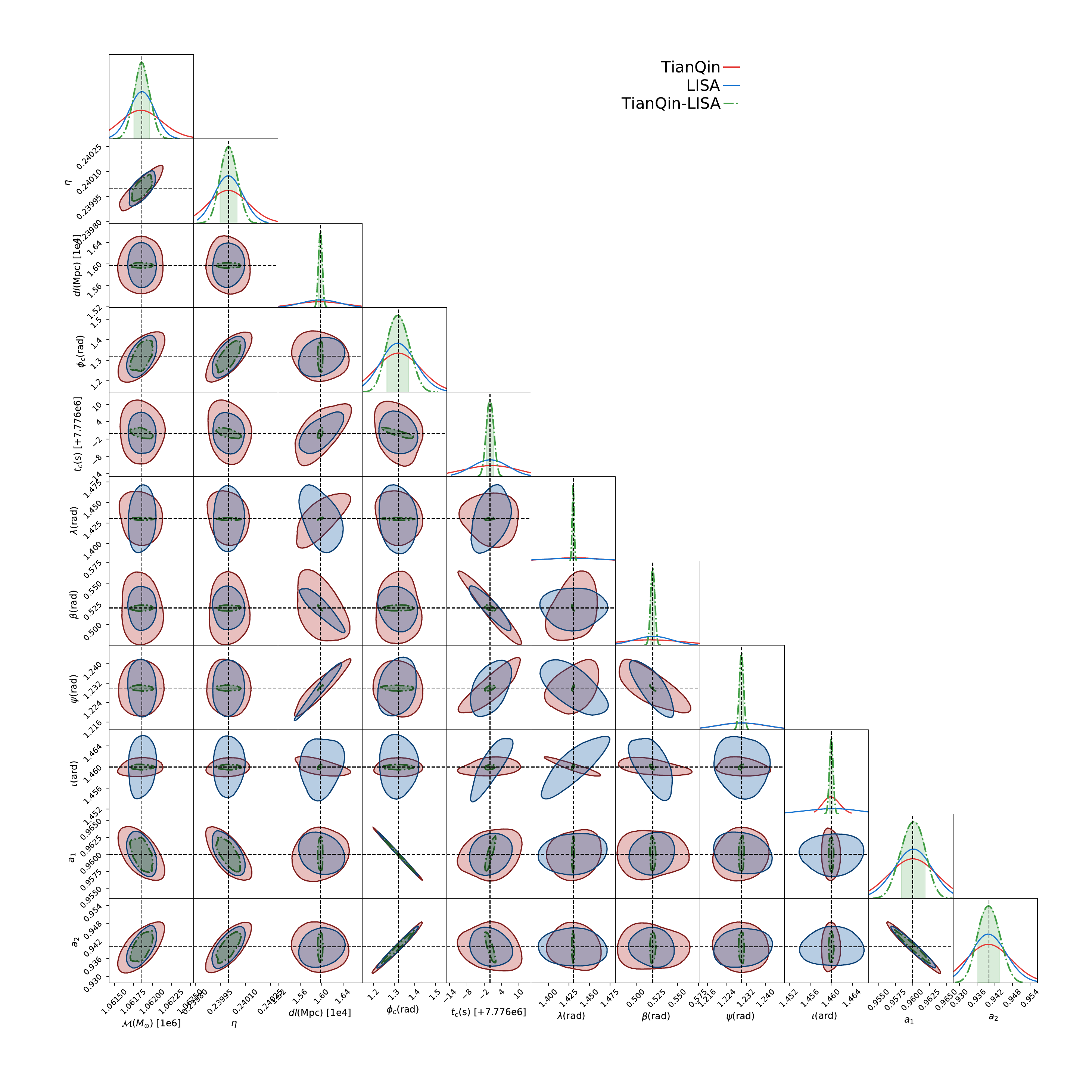}
    \caption{The 11-dimensional parameters' posterior distribution and contour plots of the F0 system with different detectors are shown above, where the red, blue, and green lines are for TianQin, LISA, and the network of TianQin-LISA. 
    The lines of the corresponding color represent one-dimensional 1$\sigma$ C.L. for each parameter and the black dashed lines indicate the injected values.}
    \label{fig:Fiducial}
\end{figure*}

\subsection{Parameter estimation}
\label{subsec:PEresults}

Using the above methods, we have successfully carried out parameter estimation within the 11-dimensional parameter space for all the example sources across three detector configurations: TianQin, LISA, and the network of TianQin and LISA.
All the constraint results are listed in Table~\ref{tb:error_emcee}, where $\Delta \bftheta = \bftheta_{\rm mean} - \bftheta_{\rm inject}$ denotes the disparity between the estimated values and the injected values.
For the convenience of comparison, we have employed relative errors for certain parameters, such as $\Delta \Mc/\Mc$ and $\Delta D_L / D_L$.

\subsubsection{Different parameter's effect on the parameter estimation}


\begin{figure*}
    \centering
    \includegraphics[width=0.97\linewidth]{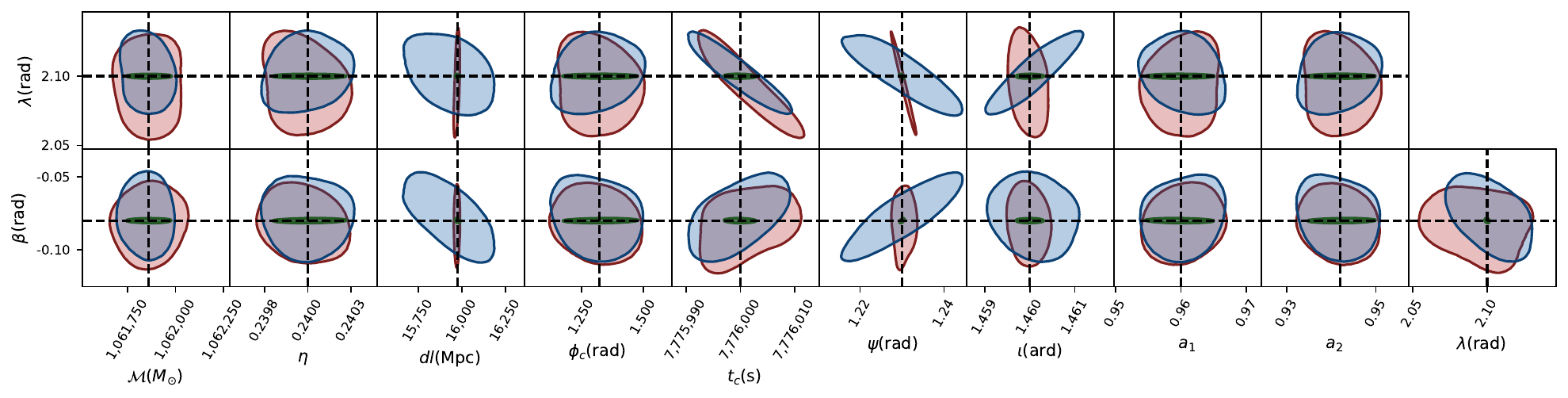}
    \caption{The contour plots of $(\lambda, \beta)$ corresponding to the other 9 parameters in FP case.
    The colors and line styles in the figure are defined following the same conventions as in Fig.~\ref{fig:Fiducial}.
    }
    \label{fig:J0806}
\end{figure*}

\begin{figure*}
  \centering
  \includegraphics[width=0.97\linewidth]{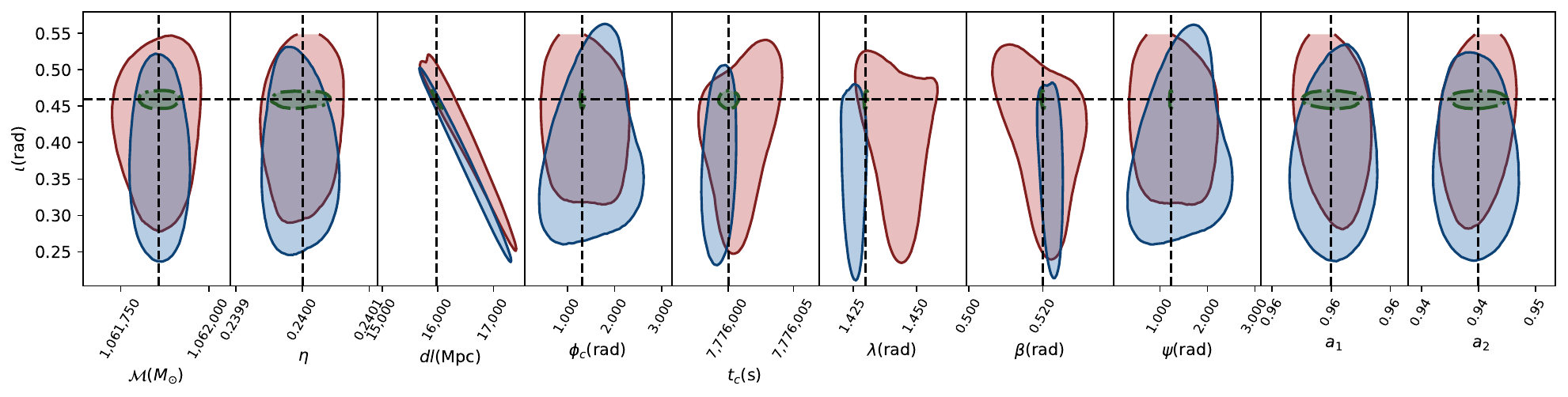}
  \caption{The contour plots of $\iota$ corresponding to the other parameters in AF case.
    The colors and line styles in the figure are defined following the same conventions as in Fig.~\ref{fig:Fiducial}.
    }
  \label{fig:Iota0}
\end{figure*}

\begin{figure*}
  \centering
  \includegraphics[width=0.97\linewidth]{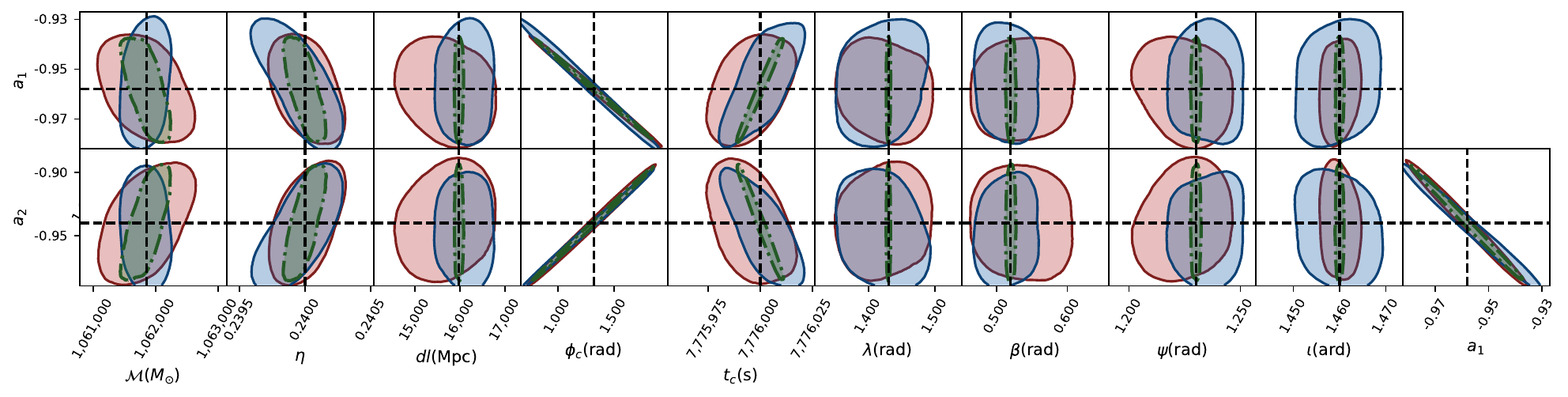}
    \caption{The contour plots of $(a_1, a_2)$ corresponding to the other parameters in AS case.
    The colors and line styles in the figure are defined following the same conventions as in Fig.~\ref{fig:Fiducial}.
    }
  \label{fig:Antispin}
\end{figure*}

\begin{figure*}
  \centering
  \includegraphics[width=0.97\linewidth]{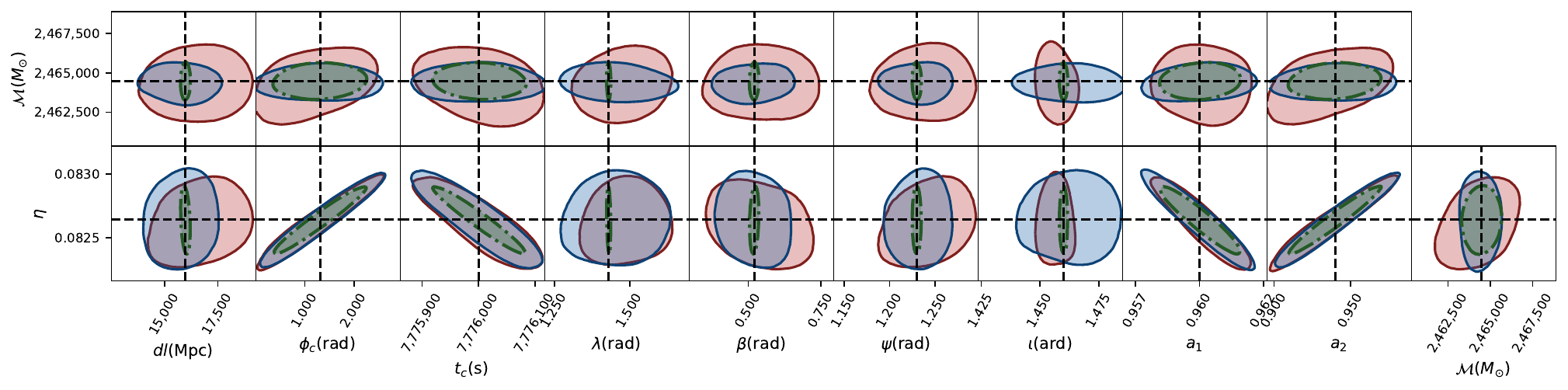}
    \caption{The contour plots of $(\Mc, \eta)$ corresponding to the other parameters in HM case. 
    The colors and line styles in the figure are defined following the same conventions as in Fig.~\ref{fig:Fiducial}. }
  \label{fig:Largemr}
\end{figure*}

\begin{figure*}
  \centering
  \includegraphics[width=0.97\linewidth]{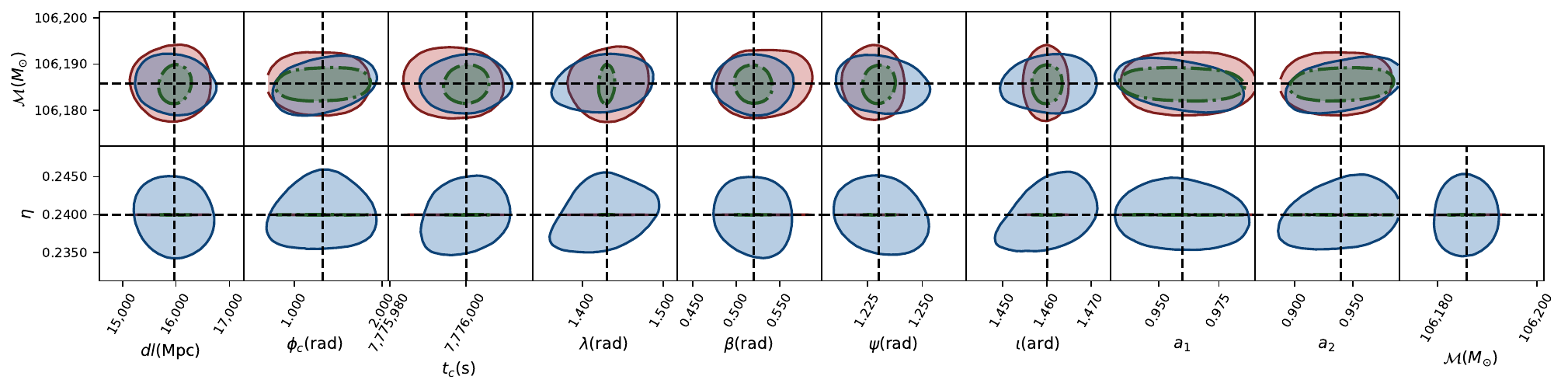}
    \caption{The contour plots of $(\Mc, \eta)$ corresponding to the other parameters in LM case. 
    The colors and line styles in the figure are defined following the same conventions as in Fig.~\ref{fig:Fiducial}. }
  \label{fig:Light}
\end{figure*}

\paragraph{The F0 case:}
Figure \ref{fig:Fiducial} displays the corner plot\footnote{The contour plots are generated using the plotting utilities \texttt{ChainConsumer} \cite{Hinton2016}.}, of the full 11-dimensional posterior distribution of the parameters for F0 case. 
The constraint results listed in Table \ref{tb:error_emcee} and the contours shown in Fig.~\ref{fig:Fiducial} reveal that the TianQin constraints are slightly looser than those of LISA, except for $\iota$ due to the smaller \ac{snr} of the source in TianQin compared to LISA (as indicated in Table \ref{params1}).

The contour plots in Fig.~\ref{fig:Fiducial} show that the degeneracy directions for several different parameter pairs differ between TianQin and LISA.
For instance, $D_L-\lambda$ and $\psi - \lambda$ exhibit a positive correlation in TianQin, while they display a negative correlation in LISA. 
Similarly, $\iota - \lambda$ shows a negative correlation in TianQin but a positive correlation in LISA. 
The joint observations of these two detectors effectively break some parameter degeneracies, leading to a significant improvement in the precision of parameter constraints.

It is worth noting that the phenomenon of degeneracy breaking mostly occurs in extrinsic parameters, such as $D_L$, $t_c$, $\lambda$, $\beta$, $\psi$, and $\iota$. 
On the contrary, the intrinsic parameters of F0 have little improvement in the constraint results with the network of TianQin and LISA. 
The parameter estimation error of some extrinsic parameters with the network has been greatly improved compared to the single detector. 
All the precision of sky position parameters $(\lambda, \beta)$ and the luminosity distance $D_L$ has improved by an order of magnitude in the network case.

Furthermore, there are strong degeneracies between $a_1 - \phi_c$, $a_2 -\phi_c$, and $a_1 - a_2$, relatively weak degeneracy between the pairs of intrinsic parameters $(\Mc, \eta, a_1, a_2)$, and the degeneracies direction is the same in TianQin and LISA. 
The coalescence frequency at which these reference values ($\phi_c$, $t_c$) are set is determined, in part, by the symmetric combination of the mass-weighted spins. 
Therefore, the network can slightly improve the constraint ability of these parameters.

Finally, the reference frequency, $f_{\rm ref}$, is set by the symmetric combination of the mass-weighted spins. 
As this binary has similar values for $a_1$ and $a_2$, the injection has a larger mass-weighted contribution from its primary.
For this reason, we see the likelihood is very sensitive to the changes in the spins when the mass-weighted spin contribution from the secondary is larger than that from the primary.

\paragraph{The FP case:}
This system's sky location coincides with that of the pole, indicating its optimal positioning for TianQin.
The two-dimensional contour plots illustrating the posterior distributions of various parameters in the FP case are presented in Figure \ref{fig:J0806}. 
As expected, the constraints on $(\lambda, \beta)$ for TianQin are significantly tighter compared to the F0 case.
In particular, the 1$\sigma$ error for the luminosity distance $D_L$ in this system is tens of times smaller than that of the F0 system. 
Additionally, there is a notable improvement by a factor of several times in the constraints for the polarization angle $\psi$ and the inclination angle $\iota$.
This improvement can be attributed to the degeneracies between $(\lambda, \beta)$ and other parameters. 
As the constraints on $(\lambda, \beta)$ improve, the constraints on the other parameters also benefit.

Compared with Figure \ref{fig:Fiducial}, the contour plot of the system reveals the disappearance of degeneracies in $D_L -\lambda$, $D_L - \beta$, and $\psi - \beta$ for the FP case in TianQin, which were present in the F0 case. 
The relationship between $\psi$ and $\lambda$ in TianQin has changed from positive to negative when compared to the F0 case. 
The degeneracy relationships of other parameters remain largely unaffected.
It is worth noting that the network still improves the precision of the sky position parameters by a factor of several times as a result of breaking the corresponding degeneracies compared with the single detector.

\paragraph{The AF case:}
In this scenario, a smaller value of $\iota$ was chosen, enabling the detectors to capture more information from the $+$ and $\times$ modes of \acp{gw}.
As a result of the increased \ac{snr}, the constraint results for almost all intrinsic parameters have become tighter compared to the F0 case. 
However, the constraints on the extrinsic parameters have deteriorated, leading to greater bias and larger errors, except $t_c$.

Figure \ref{fig:Iota0} displays the contour plots of different parameters corresponding to $\iota$. 
Upon comparison with the contour plots in Fig.~\ref{fig:Fiducial}, it becomes evident that some parameter degeneracies have vanished, except for the $D_L - \iota$ degeneracy, which has become more pronounced.

The $D_L - \iota$ degeneracy can be explained by considering that the amplitude of \ac{gw} is inversely proportional to $D_L$. 
As $\iota$ approaches $\pi/2$, the contribution of $\iota$ to the amplitude of \ac{gw} diminishes, resulting in a weak degeneracy between $D_L$ and $\iota$. 
Thus, for small values of $\iota$, the degeneracy becomes stronger.

Furthermore, the contour plot of $\lambda - \iota$ in Fig.~\ref{fig:Iota0} reveals minimal overlap between TianQin and LISA within the 1$\sigma$ region of the constraint results for $\lambda$. 
Nevertheless, when the data from both detectors are jointly analyzed, the constraint result for $\lambda$ improves significantly. 
This observation indicates that utilizing data from different detectors enhances the estimation capabilities of gravitational wave parameters.

\paragraph{The AS case:}
Contour plots of 9 model parameters are depicted in Fig.~\ref{fig:Antispin} against $(a_1, a_2)$. 
A comparison between the results obtained under the F0 and AS scenarios indicates a more loose constraint on the latter. 
Specifically, the errors observed in the $(a_1, a_2)$ parameters have increased by an order of magnitude.

Furthermore, the parameter degeneracy pattern exhibited by the individual TianQin or LISA detectors and the network remains similar to that observed under the F0 scenario. 
Notably, there is no discernible breaking of the parameter degeneracy pattern with changes to spin parameters. 
Instead, the chosen of anti-spin serves to reinforce the degeneracy existing between $(a_1, a_2)$ and $\phi_c$.

\paragraph{The HM case:}
Figure~\ref{fig:Largemr} illustrates the contour plots between $(\Mc, \eta)$ and the other parameters. 
In this case, the estimation precision of the model parameters experiences a significant decrease by several orders of magnitude compared to the F0 case, except for $a_1$, which demonstrates an improvement in estimation precision by approximately an order of magnitude.
This can be explained as follows: as the mass ratio increases, the inclusion of $a_i$ introduces a more substantial mass-weighted contribution from the primary \ac{mbh}. 
Thus, the constraint on $a_1$ becomes significantly tighter than that on $a_2$. 
This phenomenon remains consistent across all cases.

Unlike the F0 case, in the HM case, there are noticeable changes in the strength of degeneracy among $(\Mc, \eta, a_1, a_2)$. The degeneracy between $\Mc$ and $a_1, a_2$ weakens, while the degeneracy between $\eta$ and $a_1, a_2$ strengthens. 
There is virtually no remaining degeneracy observed between $\Mc$ and $\eta$.

\paragraph{The LM case:}
Figure \ref{fig:Light} displays the contour plots between $(\Mc, \eta)$ and other parameters. 
In comparison with the F0 case, the LM case demonstrates improved constraints on $\Mc$ for different detectors.
However, the precision of the spin parameter constraints has decreased. 
The precision of other parameters remains largely unchanged. 
Nonetheless, there is a difference in the performance of TianQin and LISA: TianQin improves the precision of constraint on $\eta$, whereas LISA shows the opposite trend. 
This phenomenon is attributable to the fact that as the mass decreases, the degeneracy between $\Mc -\eta$ weakens.

Moreover, compared to all other cases, the LM and FP cases demonstrate minimal improvement in the network of TianQin and LISA on the constraint of luminosity distance. 
Conversely, in the remaining cases, the constraints on $D_L$ exhibit an improvement of approximately one order.

\begin{figure*}
  \centering
  \includegraphics[width=\linewidth, trim=2.8cm 2.8cm 2.4cm 2.4cm, clip]{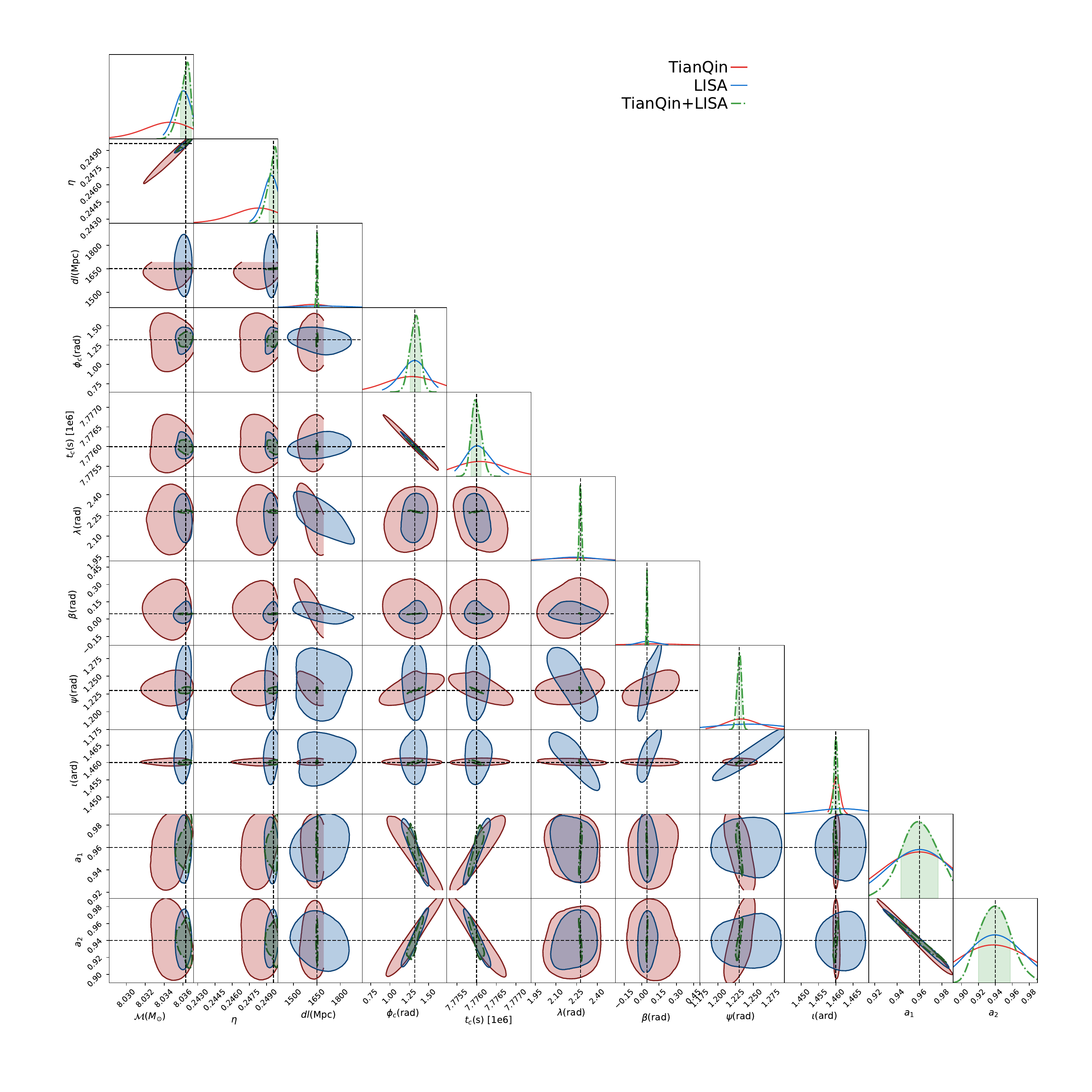}
    \caption{The contour plots of 9 parameters corresponding to $(\Mc, \eta)$ of OJ287.
    The red, blue, and green lines correspond to TianQin, LISA, and TianQin-LISA.
    The lines of the corresponding colors represent 1$\sigma$ C.L. for each parameter and the black dashed lines indicate the true parameters.}
  \label{fig:OJ287}
\end{figure*}

\begin{figure*}
  \centering
    \includegraphics[width=\linewidth, trim=2.8cm 2.8cm 2.4cm 2.4cm, clip]{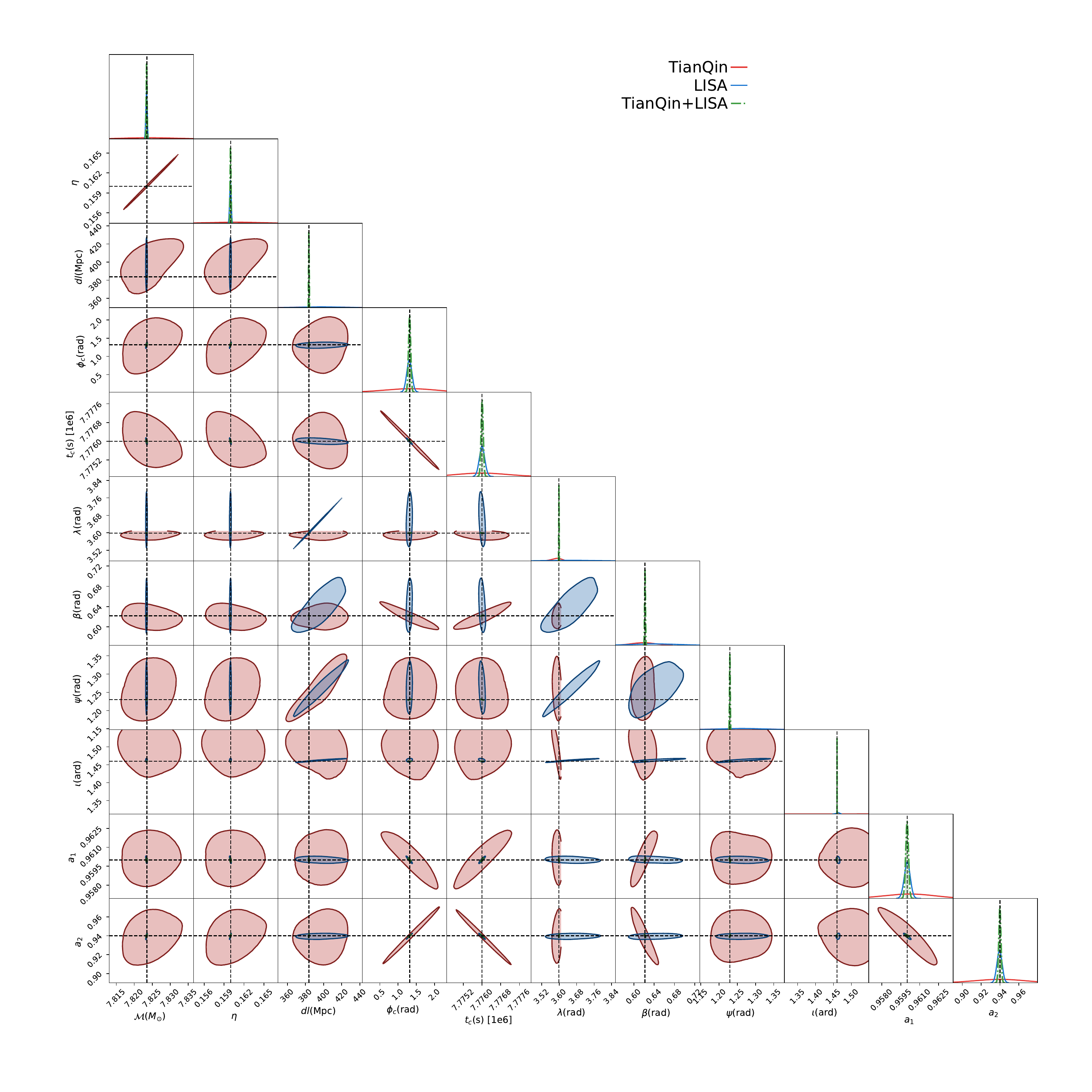}
    \caption{The contour plots of 9 parameters corresponding to $(\Mc, \eta)$ of Tick-Tock.
    The red, blue, and green lines correspond to TianQin, LISA, and TianQin-LISA.
    The lines of the corresponding color represent 1$\sigma$ C.L. for each parameter and the black dashed lines indicate the true parameters.}
  \label{fig:Ticktock}
\end{figure*}

\subsubsection{Parameter estimation of OJ287 and Tick-Tock}

In this subsection, we present the constraint results for the OJ287 and Tick-Tock systems, which is presented in Figure~\ref{fig:OJ287} and Figure~\ref{fig:Ticktock}, respectively.
As depicted in Fig.~\ref{fig:signal}, only a portion of the merger and ringdown phases of these two systems can be detected by TianQin and LISA. 
Thus, this limited detection may have a significant impact on parameter estimation.

For the OJ287 system, it can be observed that TianQin's constraint results are slightly looser compared to those of LISA, except for $(D_L, \psi, \iota)$. 
As for the Tick-Tock system, TianQin's constraints for all 11 parameters are looser than those of LISA.
This outcome aligns with expectations, as LISA is more sensitive to heavier \acp{mbhb}, and the \acp{snr} of the two systems in TianQin are considerably smaller than those in LISA (see Fig.~\ref{params2}).

Compared to the F0 case, noticeable degeneracies exist between $t_c - \phi_c$, $t_c-a_q$, and $t_c-a_2$ in the two systems. 
Additionally, the constraints on the sky location parameters $(\lambda, \beta)$ exhibit bias in both TianQin and LISA, even when considering the network of detectors. 
In contrast to the six example \ac{mbhb} systems, where there is considerably less bias in constraining $(\lambda, \beta)$, this finding indicates that information from the inspiral phase of the waveform is valuable in determining the sky location. 
This trend is also evident when comparing the HM case with other cases, where TianQin and LISA observe a smaller portion of the inspiral phase, resulting in larger biases in their constraint results for $\lambda$.

\subsubsection{Effects of noise on the parameter estimation}

\begin{figure*}
  \centering
    \includegraphics[width=\linewidth, trim=2.8cm 2.8cm 2.4cm 2.4cm, clip]{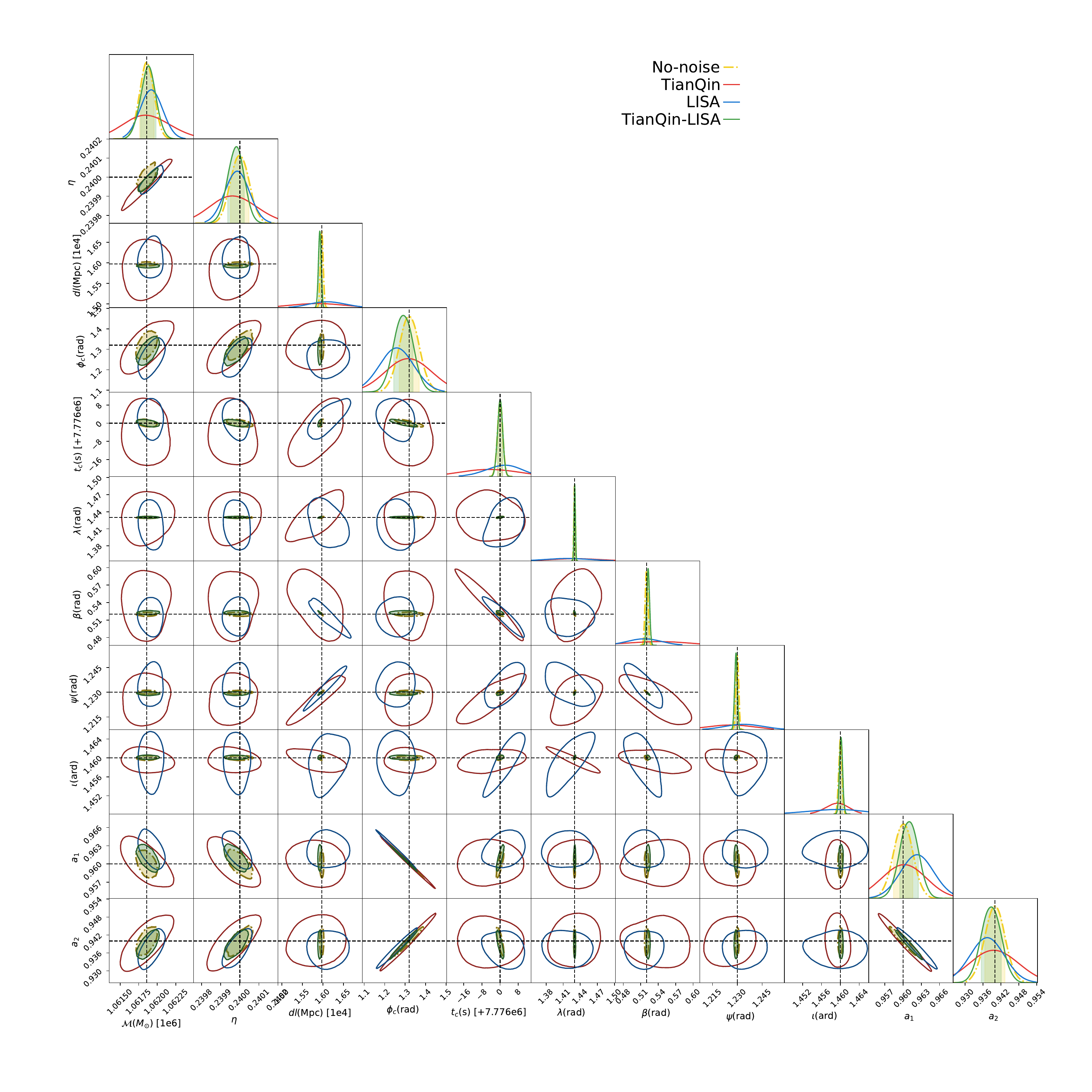}
    \caption{The contour plots of F0 case with random noise data.
    The red, blue, and green lines correspond to TianQin, LISA, and TianQin-LISA.
    The lines of the corresponding color represent 1$\sigma$ C.L. for each parameter and the black dashed lines indicate the true parameters.
    As a comparison, the constraint results of the F0 case without noise under the joint detection are also shown in the figure.}
  \label{fig:F0Noise}
\end{figure*}

To examine the impact of noisy data on the constraint results, we conducted a comparison between the F0 case with and without injected noise in the data. 
The constraint results can be found in Table~\ref{tb:error_emcee}, while the corresponding contour plots are presented in Fig.~\ref{fig:F0Noise}.

According to the contour plots, it is clear that the degeneracies between different parameters remain consistent with those observed in the F0 case. 
This indicates that the presence of stochastic noise does not change the inherent degeneracies between parameters. 
However, when noise is injected into the data, the constraint results for all parameters in either TianQin or LISA become worse. 
Most parameters exhibit biased best-fit values, and the overall error increases. 
Nevertheless, when considering the joint observations from TianQin and LISA, the constraint results closely resemble those obtained in the noise-free F0 case.

It is important to note that although the inclusion of noise in the data leads to worsened constraint results, the deviations observed are not substantial. 
It appears that all errors remain within acceptable bounds and do not exceed 2$\sigma$ level. 
Therefore, one can assert that the conclusions drawn from the analysis hold true even in the presence of noise.

\section{Discussion and conclusion}
\label{sec:conc}

In this paper, we have examined the impact of various parameter changes on the estimation of parameters for \ac{mbhb} systems.
To assess these effects, several designed \ac{mbhb} systems are chosen for comparison with a fiducial system, referred to as the F0 case, where the parameters are arbitrarily chosen. 
For a \ac{mbhb}, the masses and spins play a crucial role in determining the system's evolution.
Therefore, we have considered cases such as anti-spin (AS), large mass ratio (HM), and light mass (LM) to explore their influence.
Considering that detectors have preferential directions in response strength, we have also explored an optimal direction case for TianQin, referred to as the FP case.
The inclination angle can affect whether the detector can detect the total $+$ and $\times$ modes of the \ac{gw}.
Hence, we have considered an almost face-on case, denoted as the AF case.
The parameter values for these systems are listed in Table~\ref{params1}.
To evaluate the effectiveness of our method in detecting real MBHBs, we have considered two popular systems, OJ287 and Tick-Tock, as shown in Table~\ref{params2}.
Lastly, we have examined the effect of noise on parameter estimation.

A Bayesian inference technique is adopted to extract marginalized posterior distributions across the 11-dimensional parameter space.
The constraint results, including the recovered parameter values and their 1$\sigma$ errors for all 11 parameters under the different cases, are presented in Table~\ref{tb:error_emcee}.
The corresponding contour plots can be found in Figs.~\ref{fig:Fiducial}-\ref{fig:F0Noise}
The contour plots demonstrate that the network of TianQin and LISA can break certain degeneracies among different parameters, such as $\lambda -D_L$, $\lambda- \psi$, $\lambda-\iota$, and $\iota -\beta$, among others.
Furthermore, joint detection can improve the estimation of parameters, particularly for extrinsic parameters such as $\lambda$, $\beta$, and $\psi$, with their error reduced by approximately one order magnitude in the joint detection case.

Our constraint results reveal that the errors of most parameters do not simply follow a proportional relationship with the \ac{snr}, and the degeneracies between different parameters are significantly influenced by the parameter values.
For the system that has similar masses for the two \acp{mbh}, we observe an increase in the error of $\Mc$ as the \ac{snr} increases while the other parameters change.
By examining Fig.~\ref{fig:signal}, it is apparent that the observation length of the inspiral phase in the data has a significant effect on the errors of the symmetric mass ratio $\eta$.
Specifically, the LM case exhibits the lowest errors for $\eta$.
The precision of constraining the luminosity distance $D_L$ increases by approximately two orders of magnitude in the FP case compared to other cases, suggesting that the detector's response greatly influences the constraint on $D_L$. 
However, in the other cases, there is little improvement in the errors.
Regarding $\phi_c$, we observe larger errors in the HM and Tick-Tock cases, indicating that a higher mass ratio can lead to poorer estimation of $\phi_c$.
The anti-spin, lower mass, and reduced detection of the inspiral phase result in poorer constraints on $(a_1, a_2)$.
In addition, comparing all the systems studied, one can also find that the small inclination angles (AF case) and the limited detection of the inspiral phase (HM, OJ287, and Tick-Tock cases) can introduce significant bias in the estimation of parameters.

Generally, the network of TianQin and LISA network helps break certain degeneracies among parameters of \ac{mbhb}, especially for the extrinsic parameters.
While the network can significantly improve the estimation of certain parameters, such as sky localization, its efficacy may not extend uniformly to all parameters.
Moreover, it is important to note that even with a sufficiently high \ac{snr}, parameter estimation can still be subject to biases if the detected signal does not encompass all stages of the inspiral, merger, and ringdown.

\acknowledgements{%
J. Gao thanks Xiangyu Lyu for his help in the discussion about the response function used in TianQin.
This work has been supported by the Guangdong Major Project of Basic and Applied Basic Research (Grant No. 2019B030302001), the Natural Science Foundation of China (Grant No. 12173104, No. 12261131504), and the Natural Science Foundation of Guangdong Province of China (Grant No. 2022A1515011862).}

\bibliography{ref}

\end{document}